\documentclass{WileyMSP-template}

\usepackage{caption}
\usepackage{subcaption}
\usepackage{xcolor}

\begin{document}

\pagestyle{fancy}
\rhead{\includegraphics[width=2.5cm]{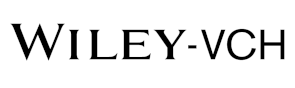}}

\title{Recent advances on quantum key distribution overcoming the linear secret key capacity bound}

\maketitle


\author{Yingqiu Mao}
\author{Pei Zeng}
\author{Teng-Yun Chen}


\begin{affiliations}
Dr. Y. Mao, Prof. T.-Y. Chen\\
Hefei National Laboratory for Physical Sciences at the Microscale and Department of Modern Physics, University of Science and Technology of China, Hefei 230026, China\\
CAS Center for Excellence and Synergetic Innovation Center in Quantum Information and Quantum Physics,
University of Science and Technology of China, Hefei, Anhui 230026, P. R. China\\

P. Zeng\\
Center for Quantum Information, Institute for Interdisciplinary Information Sciences, Tsinghua University, Beijing 100084, China\\

\end{affiliations}



\begin{abstract}

A crucial goal for quantum key distribution (QKD) is to transmit unconditionally secure keys over long distances. Previous studies show that the key rate of point-to-point QKD is limited by a secret key rate capacity bound, and higher key rates would require quantum repeaters. In $2018$, the seminal twin-field (TF) QKD protocol was proposed to provide a remarkable solution to overcoming the linear secret key capacity bound. This article presents an up-to-date survey on recent developments in this area, including the security proofs of phase-matching QKD and other TF-QKD type protocols, the theoretical examinations of these protocols under realistic conditions, and the recent experimental demonstrations.

\end{abstract}



\section{Introduction}

Secure communication is vital for protecting both personal information and national secrets in today's interconnected world. 
One of the most effective tools to obtain security is through encryption, which heavily relies on the secrecy of the keys. 
Based on the laws of quantum mechanics, quantum key distribution (QKD) provides a way to generate information-theoretic secure keys for authorized communication parties \cite{Scarani2009RMP}, making it of great practical value for network security. 
Since the first QKD protocol BB84 was introduced \cite{BB84}, numerous demonstrations have been performed in both optical fibers and free space, and commercial QKD systems have also been successfully developed \cite{xu2019secure,pirandola2019advances}. 
As one of the most mature technologies in quantum information, researchers are advancing the development of QKD to approach real-life applications \cite{xu2019secure}.

The transmission distance and secure key rate are two of the most important parameters for practical QKD. In terms of distance in optical fiber, the longest BB84 protocol demonstration has reached $421$ km \cite{boaron18}, and if side-channel attacks are considered, the distance can still reach $404$ km using the measurement-device-independent (MDI) QKD protocol \cite{yin2016mdi404,braunstein12,mdiqkd}, though in both cases the key rate is below $1$ bits/s (bps). In the meantime, field fiber tests have reached $130$ km with final key rate of $0.2$ kbps \cite{chen2010metropolitan}, while satellite-to-ground free space QKD have been achieved over $1200$ km \cite{Liao2017nature}. In terms of key rate, state-of-the-art performance have reached $>10$ Mbps over $10$ km fiber \cite{yuan201810} and $>1$ Mbps over $50$ km \cite{dynes2016ultra} with a $1$ GHz QKD system. One can also obtain high key rates at short transmission distances with continuous-variable QKD, though its security analysis for finite data is not yet as mature as that for discrete-variable QKD \cite{Xu2015}.   

To increase the performance of QKD, upgrades are necessary in for example single-photon detectors and post-processing hardware. 
However, even if this complex task is eventually accomplished, optical signals attenuate exponentially with transmission loss, and the optimal key rate of a point-to-point QKD protocol is limited by certain bounds \cite{pirandolaPRL09,tgw14,Pirandola2017}. In particular, the fundamental bound is the linear secret key capacity bound, which is also known as the PLOB bound \cite{Pirandola2017}.
For instance, with a state-of-the-art QKD system that operates the standard decoy state BB84 protocol, even using near perfect detection efficiency single-photon detectors, the longest transmission distance would only reach $\sim600$ km, where the key rate is extremely low and unsuitable for real-time applications. 

To overcome the PLOB bound and increase the transmission distance, a natural solution is to use quantum repeaters \cite{memory2011RMP}. Unfortunately, current quantum repeaters are still a long way from practical applications, so as an alternative, one may use ``trusted relays'', as shown in several QKD network demonstrations, where intermediate nodes are authorized and secured physically to relay keys between indirectly connected users \cite{peev2009secoqc,sasaki2011field,Zhang:18}. However, the trustworthiness of such relays may not be guaranteed in strict scenarios, particularly against eavesdroppers with power bounded only by quantum mechanics \cite{Qi07,Lydersen2010,Xu_2010njp,Lo2014}. This issue can be resolved with MDI-QKD, whose introduction opened a new perspective for QKD protocols \cite{braunstein12,mdiqkd}. Instead of transmitter Alice sending photons to receiver Bob, the two parties send their signals to an untrusted third party Charlie, which removes assumptions on the detection device and creates an untrusted relay between Alice and Bob. Still, as coincidence detection is required at the measurement site, it was shown that this type of protocol cannot beat the PLOB bound \cite{Pirandola2017}. Thus, finding a protocol that can beat this bound is essential for the development of practical QKD. 

In $2018$, Lucamarini \textit{et al.} proposed a new QKD protocol, i.e., twin-field (TF) QKD, that provides a remarkable solution \cite{Lucamarini2018}. The TF-QKD protocol is an efficient version of MDI-QKD, namely, it is not only immune to side-channel attacks, but can also overcome the PLOB bound. Since its introduction, intense research has been dedicated in several areas. To understand the security basis of TF-QKD, studies has been conducted to develop rigorous security proofs \cite{ma18,tamaki2018information,snsqkd,cui19,Curty2019,lin18}. Also, to accelerate the application of TF-QKD, numerous works have investigated its feasibility in practical conditions such as finite key analysis, asymmetric transmission distances, and so on \cite{Yu2019,Zhang:19tfqkd,lu2019tfqkd,Lu_2019,Yin2019,zhou2019PRA,Grasselli2019,Maeda2019,Yin2019SR,Yin2019a,wang2019twinfield,Grasselli2019asymm,Wang_2020,Li2019,hu2019PRA,jiang2019zigzag,lorenzo2019tight}. 
 Furthermore, efforts have been made in developing new experimental techniques to realize TF-QKD \cite{Minder2019,wang19,liu19,zhong19,Fang2020,chen509snstfqkd,zhong2020proof}. 
 
This article serves to survey recent developments in QKD overcoming the linear secret key capacity bound. Particularly, we analyze the TF-QKD protocol, as well as the differences among the family of TF-QKD type protocols. We focus on capturing the main properties of crucial works, while the details of theoretical security analysis are out of the scope of this review. In addition, we survey the state-of-the-art experimental developments for TF-QKD and discuss the challenges in practical applications. We would like to point out that the ordering of theoretical works summarized in this paper is mainly based on the time of submission or publication, as well as their relationships with one another, and does not reflect their significance.

\section{Limits on repeaterless QKD}

For point-to-point QKD protocols, they can be equivalent to Alice using a quantum channel $\mathcal{N}$ to send $n$ quantum states to Bob. At the same time, Alice and Bob are allowed all possible local operation and classical communication to exchange classical information. Under ideal conditions, i.e., perfect single-photon source, detectors, and error corrections, and considering only transmission loss, the secure key rate $R$ is proportional to the channel efficiency $\eta$. 
Initial studies used the reverse coherent information (RCI) to derive the optimal achievable rate over a lossy channel, showing the lower bound $R\left(\eta\right) \geq -\log_{2}(1-\eta)$ \cite{pirandolaPRL09}. Then, follow up studies adopted the squashed entanglement to derive the upper bound $R\left(\eta\right)\leq\log_{2}\left(\frac{1+\eta}{1-\eta}\right)$  \cite{tgw14}.
Finally, in 2015, further studies \cite{Pirandola2017} exploited the relative entropy of entanglement to show \begin{equation}
R_{\textrm{PLOB}}\left(\eta\right)\leq-\log_{2}\left(1-\eta\right),
\label{eq:PLOB}
\end{equation}
closing the gap with the RCI lower bound and therefore establishing the secret key capacity of the lossy channel to be exactly equal to $-\log_{2}(1-\eta)$. This is shown as black line in Fig. \ref{fig0:keyrate}.
For long transmission distances, $\eta<<1$, so $R\left(\eta\right)\simeq1.44\eta$. If a quantum repeater is placed between Alice and Bob, the key rate can be increased to \cite{Pirandola2019}
\begin{equation}
R_{\textrm{SRB}}\left(\eta\right)\leq-\log_{2}\left(1-\sqrt{\eta}\right),
\label{eq:SRB}
\end{equation} 
as shown as the pink solid line in Fig. \ref{fig0:keyrate}.

\section{Protocols}

Since the emergence of TF-QKD, researchers are quick to notice that it is drastically different from previous QKD protocols. While it promises ultra-long distance distribution of raw key information, is TF-QKD information-theoretically secure? Was the security analysis given in the original work of Ref. \cite{Lucamarini2018} correct? If not, how should the unconditional security of TF-QKD be proven? Are the existing methods for QKD security proofs sufficient for this task? If the security of TF-QKD in ideal conditions can indeed be proven, is this protocol still secure under practical conditions? These are the problems that should be addressed in the theoretical studies for TF-QKD. 

\subsection{Twin-field QKD }

\begin{figure}[h!]
\centering
\includegraphics[width=15cm,trim={0.2cm 0.8cm 0.1cm 0.8cm},clip]{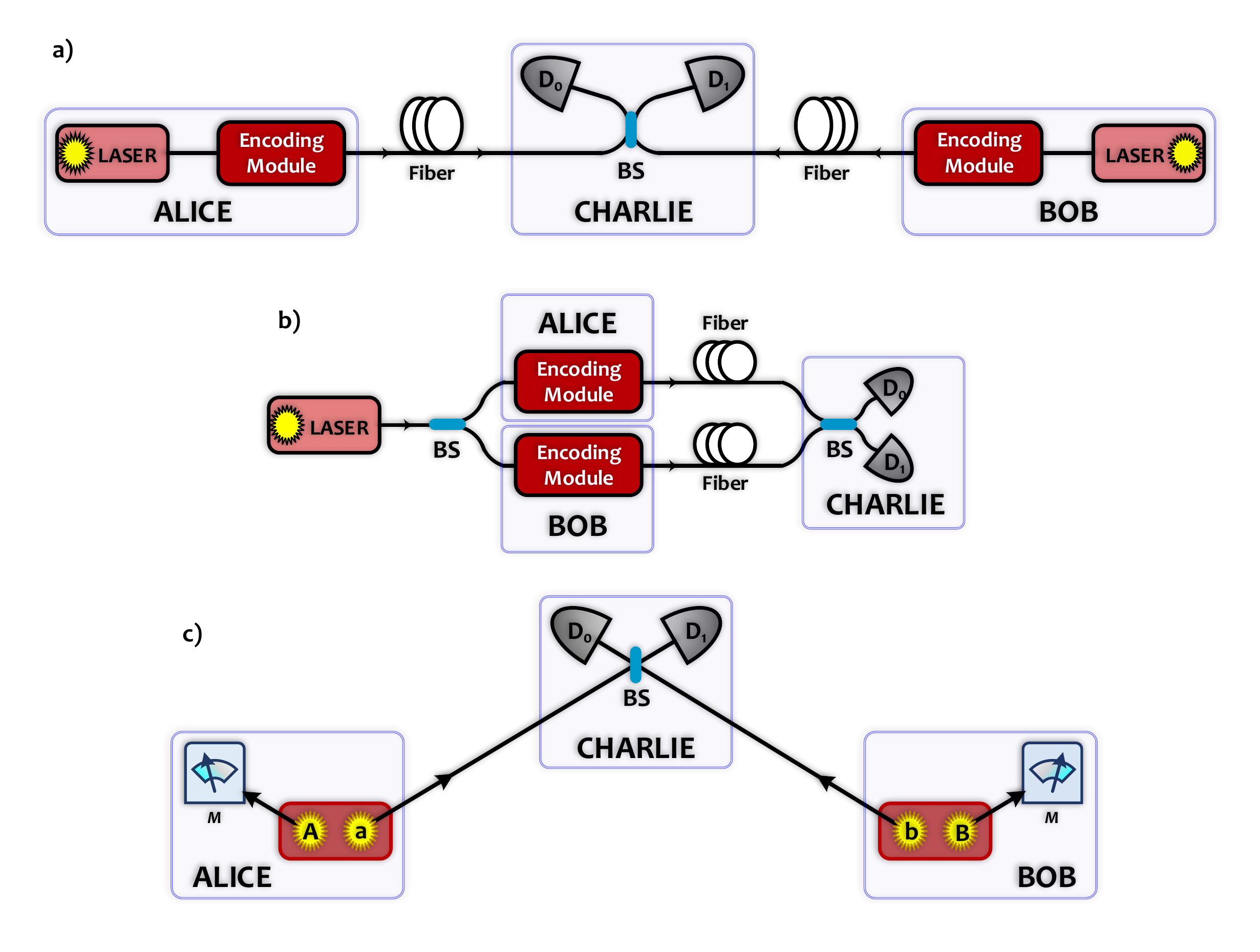}
\caption{a) Schematics of TF-QKD \cite{Lucamarini2018}. b) Interpretation of TF-QKD with a large single-photon interferometer. c) Alternative understanding for TF-QKD, where Alice and Bob each prepares an entangled state $\left|\Psi_{Aa(Bb)}\right\rangle$ and sends qubit $a$ ($b$) to Charlie for detection. When single photon interference is observed, qubits $A$ and $B$ are entangled, indicating that secure keys can be generated between Alice and Bob. BS: beam splitter; $D_0$: Detector 0; $D_1$: Detector 1; $M$: measurements in the $X$ basis or $Z$ basis.}
\label{fig1:tfqkd}
\end{figure}

The schematics of TF-QKD are shown in Fig. \ref{fig1:tfqkd}, which closely resembles phase-encoding MDI-QKD \cite{tamaki2012PRA}. However, by sending two optical fields to Charlie, when the phase of the fields satisfies ``twin'' conditions, single-photon interference results can be observed, thus enabling one to achieve $\sqrt{\eta}$ key rate scaling. In this section, we first review the procedures of TF-QKD from Ref. \cite{Lucamarini2018}, then compare it with conventional QKD protocols such as BB84 and MDI-QKD. 

\subsubsection{The procedure for TF-QKD: }
\begin{enumerate}
\item Alice and Bob separate the phase interval of $[0,2\pi)$ into $M$ phase slices, each slice has an interval $\Delta k=2\pi k/M,M=\left\{ 0,1,2,...,M-1\right\} $. 
\item Charlie is situated in the middle of Alice and Bob, i.e., let $L$ be the distance between Alice and Bob, the total transmission efficiency $\eta=10^{(-\alpha L/10)}$, then the transmission efficiency for both Alice and Bob is $\sqrt{\eta}$. By sending bright reference pulses to Charlie, Alice and Bob measure and align their phase references, where the difference mainly originates from transmission fiber induced phase fluctuations and the difference in optical source wavelengths. 
\item Alice and Bob lower their pulses to single-photon level, and encode
the following phase information onto the pulses: 
\begin{itemize}
\item Bit phase: $\alpha_{A},\alpha_{B}\in\left\{ 0,\pi\right\} $ for bits $\left\{ 0,1\right\} $, respectively; 
\item Basis phase: $\beta_{A},\beta_{B}\in\left\{ 0,\pi/2\right\} $ for $X$ and $Y$ basis, respectively; 
\item Random phase: $\rho_{A},\rho_{B}\in[0,2\pi)$, and Alice and Bob records the phase slice of their random phases,
\end{itemize}
making the total phase of the pulse $\varphi_{A},\varphi_{B}=(\alpha+\beta+\rho)\oplus2\pi$.
Also, Alice and Bob randomly adjust their pulse intensities among three values, $\mu_{A},\mu_{B}\in\left\{ \mu/2,\nu/2,\omega/2\right\} $, as the decoy states \cite{hwang2003quantum,wang2005beating,lo2005decoy}. 
\item Alice and Bob send their pulses to Charlie, who may be untrusted. Charlie interferes the pulses with a beam-splitter, and announces the results of the two detectors. Effective detections are those with only one detector clicks, this is equivalent to single-photon interference results. 
\item Basis reconciliation: 
\begin{enumerate}
\item Alice and Bob only keep effective detection events. 
\item Alice announces the decoy, basis, and phase slice of her pulses, and Bob announces his pulses that are of the same configuration. 
\item Except for the pulses prepared as $X$ basis, $\mu/2$ intensity pulses, which are used to generate the final key, Alice and Bob announce all the pulses' bit information for error estimation. If the error rate is too high, they abort the protocol.
\item Depending on Alice and Bob's bit information and the detector outcomes, Bob flips his bit if necessary. 
\end{enumerate}
\item Alice and Bob perform error correction and privacy amplification to obtain the final key. 
\end{enumerate}

\subsubsection{Deriving the secret key rate}

\paragraph{Comparison with MDI-QKD}

	Since the optical setup of TF-QKD is almost exactly the same as phase-encoding MDI-QKD \cite{tamaki2012PRA}, as shown in Fig.\ref{fig1:tfqkd}a, the most straightforward TF-QKD security proof should also be similar to MDI-QKD. However, for weak coherent state based MDI-QKD, phase randomization is required for both key states and test states \cite{ma2012}. Also, for MDI-QKD, two-photon interference is required at the untrusted measurement site, so the final key rate is proportional to $\eta$, i.e., both Alice and Bob's photons must be detected. Therefore, MDI-QKD is intrinsically different from TF-QKD, and its security proof may not be used as a reference to obtain key rate scaling that overcome the PLOB bound. It is worth mentioning one MDI-QKD protocol called MDI-B92, where Alice and Bob both transmit coherent states of $\left|\alpha\right\rangle $ and $\left|-\alpha\right\rangle $ for interference at Charlie \cite{ferenczi2013security}. Unfortunately, MDI-B92 was shown not quite robust against loss, as no test states were used, so the key rate scaling to $\sqrt{\eta}$ was not found.

\paragraph{Comparison with phase-encoding BB84}

TF-QKD can also be viewed as a variant of phase-encoding BB84 QKD \cite{townsend93}. 
Since the effective detection for TF-QKD relies on single-photon
interference results, one could imagine a large single-photon interferometer schematic in the form of Fig.~\ref{fig1:tfqkd}b, where a shared source emits pulses that are split in two to be encoded respectively by Alice and Bob. Here, one can clearly see that the gain for detection events is proportional to $\sqrt{\eta}$, which suggests the key rate to surpass the PLOB bound. The total encoding phases of Alice and Bob would have to satisfy $\Delta\varphi=\left|\varphi_{A}-\varphi_{B}\right|=0$ or $\pi$ to observe only one click between $D_0$ and $D_1$.

In Ref. \cite{Lucamarini2018}, the authors showed that this large single-photon interferometer could then be further reduced to only two parties, Alice and Bob. Here, Alice holds the source. She splits her pulses into two beams, encodes one and sends both of them to Bob via the same length of fiber, who encodes the unmodulated pulse and interferes both pulses for detection. If one further replaces the large single-photon interferometer with two asymmetric Mach-Zehnder interferometers, where Alice encodes the pulses with phase information and Bob decodes the pulses at his site, one could reduce the TF-QKD schematics to the phase-encoding BB84 protocol. 

With this revelation, Ref. \cite{Lucamarini2018} argued that the tagging method used to analyze the key rate of BB84 \cite{GLLP04}, may be applied to TF-QKD. However, as TF-QKD can also be interpreted as a large single photon interferometer, and that each effective detector click is resulted from only one photon, the authors derived that ideally the key rate would be proportional to the single photon gain $Q_{1,\mu}=\mu e^{-\mu}Y_{1}$, where the single-photon yield $Y_{1}=\sqrt{\eta}$ (here, dark counts and detector efficiency imperfections are neglected). By taking $Q_{1,\mu}$ into the GLLP analysis, i.e., \cite{Lucamarini2018}, 
\begin{equation}
R_{\textrm{TF}}\geq q\left\{ Q_{1,\mu}\left[1-H_{2}\left(e_{1}\right)\right]-f Q_{\mu}H_{2}\left(E_{\mu}\right)\right\} ,
\label{eq:TFQKD}
\end{equation}
which shows that the key rate scales to $\sqrt{\eta}$. Here, $q$ is the sifting factor, $f$ is the error correction efficiency, $H_{2}(x)=-x{\textrm{log}}_{2}(x)-(1-x){\textrm{log}}_{2}(1-x)$ is the binary entropy function, $Q_{\mu}$ and $E_{\mu}$ are the total gain and QBER, respectively, and $e_{1}$ is the QBER for single photons. For practical TF-QKD with realistic parameters such as limited number of decoys, etc., the key rate may be lower, but the scaling to $\sqrt{\eta}$ does not change. This was an incredible feature, which has never been achieved with previous QKD protocols.

\paragraph{Open questions}

As TF-QKD was shown to possibly provide higher key rates than any other discrete-variable based QKD protocol while possessing the advantages of MDI, the most urgent research task was proving the information-theoretic security of TF-QKD.
As pointed out in Ref. \cite{Lucamarini2018}, by treating TF-QKD as an alternative version of phase-encoding BB84 and deriving the key rate with the GLLP method, its security against general attacks was not analyzed. 

The problem with adopting the GLLP method originates from the difference between TF-QKD and phase-encoding BB84. With phase-encoding BB84, the global phase is never revealed, hence under the condition of phase randomization the photon number channel is established, and secure keys can be drawn from single-photon events. With TF-QKD, on the other hand, some information of the pulses' global phase will be disclosed, which makes this protocol drastically different from conventional QKD protocols and cause the prerequisites for valid GLLP analysis to not hold. Indeed, a specific question to be addressed is whether the security of TF-QKD is built on simply single-photon events. Thus, while with the GLLP method Ref. \cite{Lucamarini2018} has offered a key rate function by placing restrictions on Eve's hacking ability, whether this security analysis is rigorous under all collective attacks or coherent attacks is unclear. 

Also, although one may gain understanding TF-QKD with the large single-photon interferometer picture, it may not help to derive a security analysis from it. This is because a precondition of Alice and Bob sharing a source is supposed here, but in reality the trustworthiness of such a source cannot be guaranteed, or in some cases is an unreasonable precondition as the protocol required two independent sources. These factors should also be considered in the security analysis.

The information-theoretic security for TF-QKD was eventually proven through the TF-QKD* protocol \cite{tamaki2018information}. By phase-encoding ``Code mode'' states and adopting the decoy state method for ``Test mode'' states, the secret key rate can be written as
\begin{equation}
R_{\textrm{TF*}} \geq N_{\textrm{sif},Z} [1- H_2 (e_{ph})] -\lambda_{EC},
\label{eq:TFstar}
\end{equation}
where $N_{\textrm{sif},Z}$ is the sifted key from the $Z$ basis, $e_{ph}$ is the phase error rate, and $\lambda_{EC}$ is the amount of information leaked to Eve from information reconciliation \cite{tamaki2018information}. However, due to the complexity of the proof for TF-QKD*, the security proof for TF-QKD are usually referred to as the works described in the following sections.

\subsection{Phase matching quantum key distribution }

To prove the security of TF-QKD, one must seek approaches different from qubit-based methods. In Ref. \cite{ma18}, through introducing a new protocol called phase-matching (PM) QKD, the first rigorous security analysis for these coherent-state-based QKD protocols was given. By proving PM-QKD can surpass the PLOB bound, a solid basis for theoretical and experimental research of TF-QKD was established. 

\subsubsection{Proving its security}

First, a simplified model for TF-QKD protocol should be established. In Ref. \cite{Lucamarini2018}, TF-QKD was developed according to the BB84 bases, yet all the pulses of $Y$ basis, as well as decoys of $X$ basis are all used for parameter estimation. This causes the protocol to be resource inefficient, so the protocol should be further optimized for security analysis. In PM-QKD, the authors eliminated the basis phase encoding step for state preparation, so the coherent states are encoded as $\left|\sqrt{\mu_{A(B)}}e^{i(\alpha_{A(B)}+\rho_{A(B)})}\right\rangle $. Also, for the basis reconciliation step, only key events that satisfy $\Delta\varphi=|\varphi_{A}-\varphi_{B}|=0$ or $\pi$ are used to generate the final key, while the rest are all used for parameter estimation. 

\begin{figure}[h!]
\centering
\includegraphics[width=12cm,trim={0.2cm 0.2cm 0.1cm 0.2cm},clip]{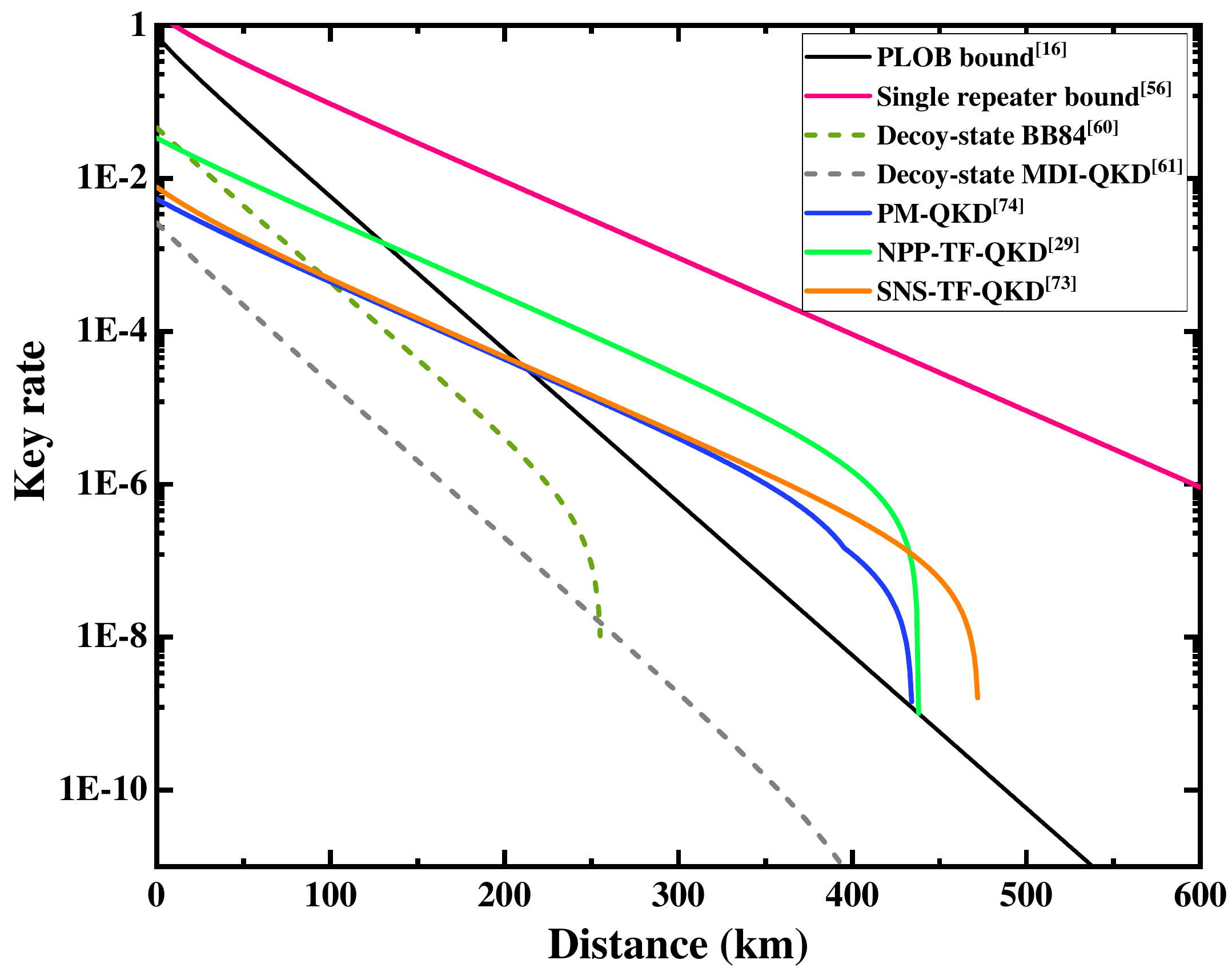}
\caption{Simulation of key rates for various protocols. Here, the error correction efficiency $f$ is $1.15$, misalignment $e_d$ is $1.5\%$, dark count rate $p_d$ is $1\times 10^{-7}$, detector efficiency is $40\%$, the number of phase slices $M$ is $16$, the average photon number $\mu$ is optimized at each distance, and the fiber loss coefficient $0.2$ dB/km. The simulation for each protocol is performed according to the respective reference in the legend. } 
\label{fig0:keyrate}
\end{figure}

To prove the security of PM-QKD, the authors began from the Lo-Chau and Shor-Preskill security argument to establish a virtual entanglement-based QKD protocol \cite{Lo2000science,shor00}. Then,  by supposing a trusted party emits a state $\rho_{AB}$ to assist Alice and Bob's local qubits for key generation, the security of an equivalent virtual protocol with coherent states was analyzed. Then, by gradually removing key preconditions such as the shared laser and BS of Fig.~\ref{fig1:tfqkd}b and phase randomization, the virtual protocol was shown to be equivalent to PM-QKD, and an optical-mode-based security proof method was developed to prove its unconditional security. An explicit key rate function for PM-QKD was shown to be in the form of \cite{ma18}
\begin{equation}
R_{\textrm{PM}}\geq \frac{2}{M} Q_{\mu}[(1-H_{2}(E_{\mu}^Z)-H_{2}(E_{\mu}^X)].
\label{eq:PMQKD}
\end{equation} 
Here, the total gain of the pulses $Q_{\mu}$ and the $Z$ basis QBER $E_{\mu}^Z$ can both be obtained directly through experiment, while phase error rate $E_{\mu}^X$ can be estimated using decoys. The key rate at different transmission distances for PM-QKD is shown as the blue solid line in Fig. \ref{fig0:keyrate}, which scales to $\sqrt{\eta}$. 

With this PM-QKD protocol, the authors further demonstrated the invalidity of the GLLP method. Using a beam splitter attack, the expected key rate obtained with the GLLP method exceeds that allowed by the attack, while the key rate of the above rigorous security analysis does not \cite{ma18}. This not only confirmed that GLLP does not work when the global phase is partially announced, but also indicated that the secure keys generated by PM-QKD come from coherent states as an entity instead of single photons. This was a remarkable revelation of PM-QKD and further underlines its intrinsic differences from previous QKD protocols. 

\subsubsection{Discussions}

For PM-QKD and MDI-QKD, the untrusted relay in the middle can act similar to a quantum repeater in that the quantum signals only needs to transmit half of the overall communication distance, so intuitively the key rate should be higher than that of point-to-point QKD protocols. Yet only by applying PM-QKD, where coherent states are used as qubit carriers and single-photon interference is invoked at the measurement site, can one overcome the PLOB bound.  
Thus, the optical-mode-based security proof in Ref. \cite{ma18} also provides a new method to investigate the use and advantages of coherent states in quantum communication. 

From a practical perspective, although PM-QKD can overcome the PLOB bound, as the authors pointed out, the key rate is still limited by a phase-sifting factor $2/M$. This is mainly due to the fact that, 
just like TF-QKD, both protocols require the random phases $\varphi_{A(B)}$ to fall in the same phase slice in order to reduce the intrinsic QBER. This way, a large amount of data is discarded from key generation. How to lower this sifting cost is still an open question. 

\subsection{Improving the protocol}

In both TF-QKD and PM-QKD, active phase randomization and phase post-selection is required, which not only makes the protocol experimentally complicated, but also causes the phase compensation to have high sifting costs. This in turn affects the security proof in two ways. First, an intrinsic sifting factor is imposed on the key rate function. Second, due to the disclosure of global phase, more complex security proof methods are required, which may cause inadequately tight bounds on error estimation and cause the key rate to be too low. To optimize the key rate and security proofs, several theoretical works were carried out around the same time, and new TF-QKD type protocols were proposed.

\subsubsection{Protocol model for analysis}

TF-QKD and PM-QKD can be generalized to the PM-MDI-QKD protocol \cite{lin18}. Here, instead of using nonorthogonal bases, two types of states are used: key states for key generation, and test states for parameter estimation. The steps for protocol include:
\begin{enumerate}
\item Alice and Bob each randomly generates a bit $m_{A}$ ($m_{B}$) to decide the state mode, where $m_{A(B)}=0$ is the key state and $m_{A(B)}=1$ is the test state. 
\item For a key mode, Alice and Bob choose phases $\alpha_{A},\alpha_{B}\in\left\{ 0,\pi\right\} $ and pulse intensities $\mu_{A},\mu_{B}$, and prepare the state as $\left|\sqrt{\mu_{A(B)}}e^{i\alpha_{A(B)}\pi}\right\rangle $. For a test mode, Alice and Bob choose $\rho_{A},\rho_{B}\in\left[0,2\pi\right)$ and prepare the state as $\left|\sqrt{\mu}e^{i\rho_{A(B)}\pi}\right\rangle $. 
\item The pulses are sent to Charlie for joint measurement, where he announces the detection possibilities: left click, right click, no click, double click. 
\item Alice and Bob disclose their choices of $m_{A}$ and $m_{B}$. Only pulse pairs with $m_{A}=m_{B}=0$ and that only one detector clicked are used for key generation. All other sets are used for parameter estimation. 
\item By random sampling the remaining bits and announcing the choices of $\mu_{A},\mu_{B},\alpha_{A},\alpha_{B}$, Alice and Bob estimates how much information Eve has on the keys. If the error rate is too high, they abort the protocol.
\item Alice and Bob perform error correction and privacy amplification to obtain the final key. 
\end{enumerate}
By using different test states, such as phase-randomized coherent states or coherent states without phase-randomization, one can obtain several variations of TF-QKD.

\subsubsection{TF-QKD with no phase post-selection}

Eliminating phase post-selection would not only lower the sifting cost and the complexity of security proofs, but also re-establish the photon number channel, which would enable applications of familiar 
techniques such as the decoy method. In two independent works of Ref. \cite{cui19} and Ref. \cite{Curty2019}, Alice and Bob prepare key states as coherent states of $\left|\alpha\right\rangle $ and $\left|-\alpha\right\rangle $, and test states as phase randomized coherent states, whose mean photon number can be chosen from a set of values. This way, since Eve cannot distinguish the coherent states for key generation from those for error estimation (assuming the mode information cannot be leaked by the public communication channel), to gain the key information she will have to splits all transmitted pulses from Alice and Bob. With the decoy state method, Alice and Bob can monitor the channel for eavesdropping. In Ref. \cite{cui19}, a general collective attack was constructed to characterize Eve's information on the key bits. Then, according to Devetak-Winter's bound \cite{DWbound}, the secret key rate function was derived. We simulate the key rate as the green solid line of Fig. \ref{fig0:keyrate}.

In Ref. \cite{Curty2019}, the security proof was based on entanglement distillation protocol as well as the loss-tolerant QKD protocol \cite{losstolerant}. The idea of the proof can be understood with Fig. \ref{fig1:tfqkd}c. Alice and Bob each locally prepares an entangled state in the form of $\left|\Psi_{Aa(Bb)}\right\rangle=\sqrt{q}\left|0\right\rangle_{A(B)}\left|0\right\rangle_{a(b)}+\sqrt{1-q}\left|1\right\rangle_{A(B)}\left|1\right\rangle_{a(b)}$. Here, qubit $a$ ($b$) is sent to Charlie, with $\left|0\right\rangle_{a(b)}$ representing the vacuum state and $\left|1\right\rangle_{a(b)}$ representing the single-photon state, and qubit $A$ ($B$) is held by Alice (Bob), with $\{\left|0\right\rangle_{A(B)},\left|1\right\rangle_{A(B)}\}$ as eigenstates of the $Z$ basis. Drawing inspiration from entanglement creation for two remote quantum repeaters \cite{Duan2001}, when single-photon interference is observed at Charlie, an entangled state can be observed as $\left|\psi_{AB}\right\rangle=\frac{1}{\sqrt{2}}(\left|1\right\rangle_{A}\left|0\right\rangle_{B}+\left|0\right\rangle_{A}\left|1\right\rangle_{B})$. Then, secure keys can be generated by performing $X$ basis measurement on qubits $A$ and $B$, where the eigenstates of $X$ basis are $\{\left|+\right\rangle_{A(B)}=\frac{1}{\sqrt{2}}(\left|1\right\rangle_{A(B)}+\left|0\right\rangle_{A(B)}),\{\left|-\right\rangle_{A(B)}=\frac{1}{\sqrt{2}}(\left|1\right\rangle_{A(B)}-\left|0\right\rangle_{A(B)})\}$, while the phase error rate can be obtained by measuring the $Z$ basis. With this virtual protocol, TF-QKD can be, just as side-channel-free QKD \cite{braunstein12}, reduced to a simple entanglement-swapping setup \cite{biham1996,Inamori2002}, and can be easily extended to the coherent-state based protocol.

Another related work was proposed by Lin and Luktinhaus, where the key states are prepared the same as \cite{cui19} and \cite{Curty2019}, but test states are even more simplified by using only coherent states of specific global phases \cite{lin18}. The security against collective attacks in the infinite key scenario was proven, which is based on the tomographic results of the quantum channel and untrusted measurement devices, obtained by using the properties of test coherent states. Note that with this proof, the key rate of PM-MDI-QKD was shown to be \cite{lin18}
\begin{equation}
R_{\textrm{PM-MDI}} = (1-e^{-2\mu\sqrt{\eta}}) \Big [ 1-H_2\Big(\frac{1-e^{-4\mu(1-\sqrt{\eta})e^{-2\mu\sqrt{\eta}}}}{2}\Big)\Big]
\label{eq:pmmdi}
\end{equation}
under loss-only conditions. For long distances and optimized $\mu$, $R$ was shown to approximate $0.0714\sqrt{\eta}$, which confirms the scaling of $R=O(\sqrt{\eta})$. 

\subsubsection{Sending-or-not-sending TF-QKD }

From an experimental perspective, as transmission distance increases, random phase fluctuations in the fiber will grow stronger and faster, which will make phase feedback difficult. In this sense, TF-QKD with no phase post-selection may encounter experimental challenges in ultra long distance implementations. In Ref. \cite{snsqkd}, a different way to combat the phase compensation problem was proposed. 
The key states, which are called $Z$ basis states, are prepared as phase randomized coherent state. For Alice, sending a pulse encodes $0$, with probability $\epsilon$, and not sending encodes $1$,  with probability $1-\epsilon$. For Bob, sending a pulse encodes $1$, with probability $\epsilon$, and not sending encodes $0$,  with probability $1-\epsilon$. Hence, this protocol is called sending-or-not-sending (SNS) TF-QKD. Effective detection is still when only one detector clicks, which means that either Alice or Bob had send a pulse and no interference had occurred, and that the secure keys come from the single-photon components. For the test states, the pulses are prepared as $X$ basis states $\left|\sqrt{\mu_{A(B)}^{\prime}}e^{i\rho_{A(B)}\pi}\right\rangle $. An effective test state detection not only requires one detector click, but also that the random phases should also satisfy a condition: $1-\left|\cos\rho_{A}-\cos\rho_{B}\right|\leq\left|\lambda\right|$, where $\lambda$ is related to the size of phase slice. During this protocol, strong optical pulses are sent from Alice and Bob to Charlie to measure the global phase. We plot the simulated key rate as the orange solid line in Fig. \ref{fig0:keyrate}.

The advantage of the SNS-TF-QKD are obvious. Since the key states are encoded as ``sending'' or ``not sending'' the coherent states, interference is not required at Charlie's measurement unit, which 
makes the $Z$ basis highly robust against optical misalignments between Alice and Bob's phase and polarization references, i.e., independent of the indistinguishability between the two lasers. In this way, the errors are mainly contributed by $X$ basis states. This will be a great advantage for achieving long transmission distances, as the key information is no longer encoded in the phase of the key states. 
Under the asymptotic scenario, the performance of SNS-TF-QKD was shown to be higher than that of TF-QKD with no phase post-selection for long distances. For short and medium distances, its key rate is still lower than that achieved with no phase post-selection TF-QKD, which is mainly due to the fact that to guarantee high visibility for single-photon interference, the sending probability for both Alice and Bob in SNS-TF-QKD must be kept low. To overcome this issue, one may apply the so-called actively odd-parity pairing method \cite{aopp}, which includes error rejection through randomly pairing and parity check. With this technique, nearly equivalent key rates as TF-QKD with no phase post-selection over short and medium distances can be achieved, and the secret key rate distance can be increased by over 50 km \cite{aopp}.

\subsubsection{Considerations for practical conditions}

These early works of TF-QKD were proven under idealistic conditions such as infinite data size, Charlie situated exactly in the middle of Alice and Bob, and that Alice and Bob have perfect sources, which means that the optical intensity and phase modification can be prepared without fluctuation. However, these conditions cannot perfectly be met with practical QKD components and systems, which will in turn affect the actual key rate achieved. Hence, we need to incorporate these imperfections into security proofs and develop security analysis under practical conditions. 

\paragraph{Finite effects}

The key rates of earliest protocols were shown to beat the PLOB bounds only when infinitely large number of pulses are sent. In reality, for $\leq$GHz repetition rate QKD systems and long transmission distances of several hundred kilometers, the effective detections will be very low, so a QKD system may need to run hours, days, or even longer to collect enough data and beat the PLOB bound. Several works have discussed the finite size effect for variations of TF-type QKD protocols \cite{Yu2019,Zhang:19tfqkd,Lu_2019,Maeda2019,Yin2019,jiang2019zigzag,lorenzo2019tight,jiang2019,zeng2020}, and have confirmed that for many protocols, the key rate can surpass the PLOB bound for $\sim10^{12}$ block size. 

\paragraph{Asymmetric channels}

In reality, Charlie will most likely not be located at exactly the center of the users. If the symmetric protocol is directly applied, and the transmission losses $\eta_{A}$ and $\eta_{B}$ are typically set to the less of the two, the key rate will be much lower than the case of $\frac{\eta_{A}}{\eta_{B}}=1$. When the channels are heavily asymmetric, e.g., the difference of Alice to Charlie and Bob to Charlie is 100 km, the key rate may not be able to surpass the PLOB bound. Similar to original MDI-QKD \cite{wangPRX19,liu19mdi}, a simple solution is to add extra fibers at the shorter side or attenuate this side's pulse power to compensate the difference between $\eta_{A}$ and $\eta_{B}$, so that the ideal symmetric protocol can still be applied, but in this case the final key rate will be limited by the side with the larger loss. Several works have investigated this issue \cite{zhou2019PRA,Curty2019,Wang_2020,jiang2019zigzag}, and have found that by independently optimizing the intensities of Alice and Bob's pulses, the key rate can be drastically increased, though it may still be lower than that of the symmetric protocol. 

\paragraph{Misalignments}

Another problem is that even when the channel losses are symmetric, fluctuations in Alice and Bob's laser emission power, which may be due to imperfection of preparation devices, noise, and so on, still 
exist. This mainly will cause incorrect estimations of the detection yields, and such deviations may be exploited by Eve. We can take this issue as instead of a constant value for average photon number $\mu$, it now fluctuates as $\mu\pm\Delta_{\mu}$, which can be accurately measured experimentally. By taking the upper and lower bounds of $\mu$, we can tightly bound the detection yields and errors, and improve the key rate calculations \cite{Lu_2019}.

\paragraph{Optimizing decoys}

The security proofs of TF-QKD usually assume the use of infinite number of decoys. In practice, however, the number of decoy states is limited, and the intensities of each setting should be optimized
for each transmission distance. Several works have addressed this issue \cite{Yu2019,lu2019tfqkd,Grasselli2019}, and have shown that generally three or four decoy states are sufficient to overcome the PLOB bound. 

\section{Experimental developments}

Having proven the security of TF-QKD under ideal and practical conditions, the next step would be testing the experimental validity of TF-QKD. Experimentally speaking, TF-QKD basically relies on pulses from two independent lasers to interfere at a middle station, after traveling through long distance fibers. For this interference to be successful, it is crucial that the optical fields from Alice and Bob's lasers are made to be ``twins''. Thus, this is the main challenge in realizing TF-QKD, how closely can Alice and Bob's lasers be made ``twins'', how to compensate or monitor the phase noise caused by fibers, and how much resources needed to overcome these issues. 

\subsection{The optical source}

In this section, we discuss the source requirements for TF-QKD and ignore the phase noise from transmission. 

\subsubsection{Considerations }

From an experimental perspective, key generation in TF-QKD basically relies on the beat note, i.e., the measured intensity of imposing Alice and Bob's laser beams on a BS. For such results to be observed with high interference visibility, some conditions should be met: 1) the temporal distributions of the two light fields must overlap; 2) the polarization states must be the same; 3) the phase of both lasers is correlated. 

In past QKD experiments, internally modulated pulsed lasers, e.g., distributed feedback lasers, were usually used to encode the bit information. For this type of source, each output pulse will have a completely random phase compared to its previous pulse, which is an ideal source for realizing protocols that require phase randomization, but this will make it difficult to control the global phase of the pulse, as required by TF-QKD, which will lead to a high portion of ineffective detection events. 

Thus, in order to implement TF-QKD, modifications on the source is necessary. Take the continuous wave (CW) laser, for example. Suppose Alice and Bob each have an ideal CW laser with perfectly aligned wavelengths $\omega$, and they send the beams through exactly the same length of fibers to Charlie (for now we ignore the decoherence effects from fibers and take them as erasure channels, and that no encoding modulations are applied), the beating result would always be steady, as the beams from Alice and Bob have phases $\gamma_{A}(t)=\gamma_{0A}+\omega t$,$\gamma_{B}(t)=\gamma_{0B}+\omega t$, respectively, and the relative phase is constant $\Delta\gamma=\left|\gamma_{0A}-\gamma_{0B}\right|$, $\gamma_{0A}$ and $\gamma_{0B}$ are the initial phases of the lasers. To encode the bit information, Alice and Bob can then use intensity modulators to slice their CW beams into pulses. In order to maintain the steady interference at Charlie, all they would need is to align their pulses to overlap precisely at the BS. This can be easily achieved by synchronizing the two lasers and adjusting the delays of the pulses. 

\begin{figure}[h!]
\centering
\includegraphics[width=12cm,trim={3.8cm 3.5cm 8.8cm 3.5cm},clip]{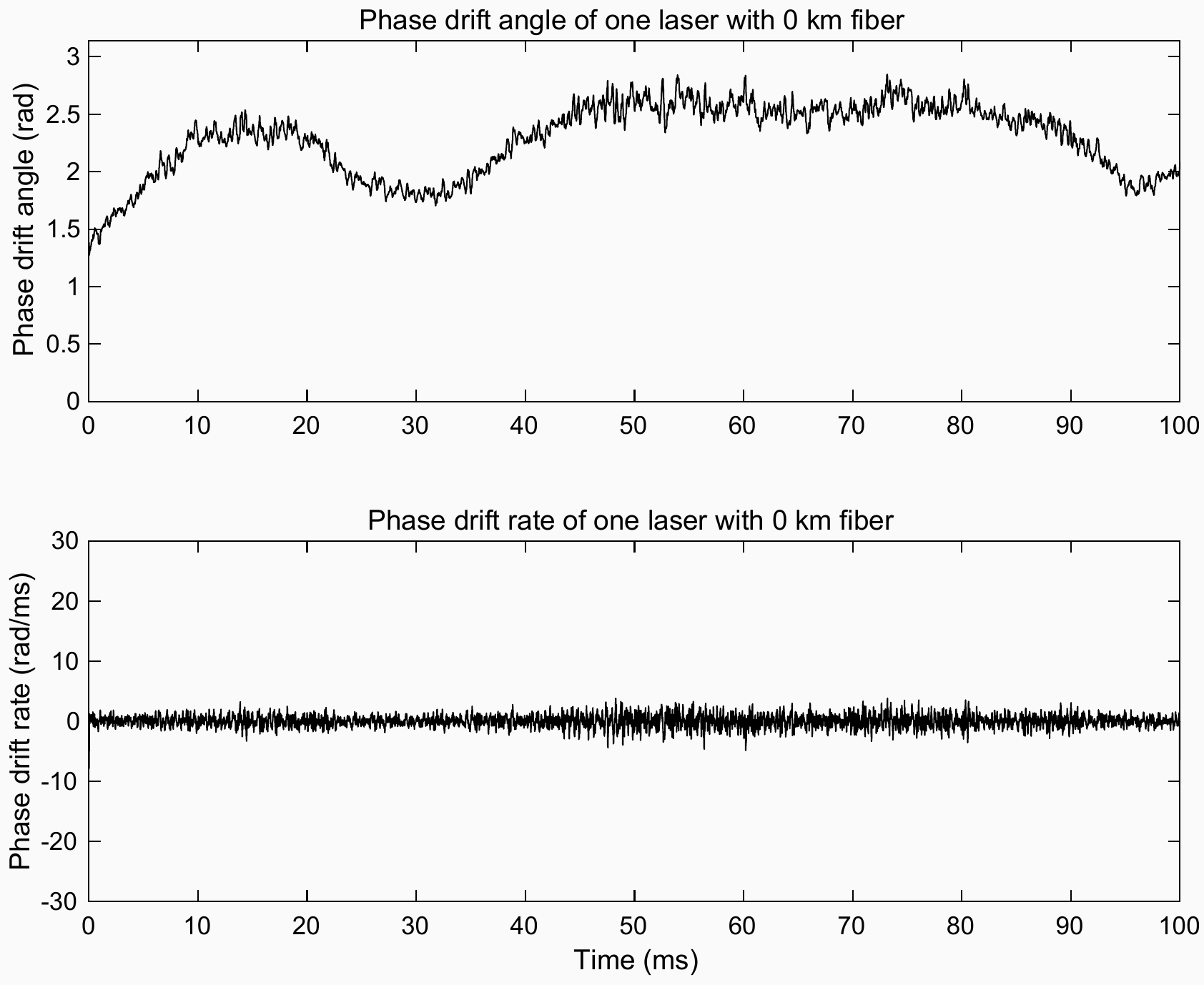}
\caption{The relative phase drift by interfering two beams split from one laser source. Reproduced with permission.\textsuperscript{[Ref. \cite{liu19}]} Copyright 2019, APS.}
\label{fig:beating}
\end{figure}

Yet these CW lasers are ideal in the way that in the frequency domain the linewidth of laser beam is a delta function, while in reality due to vibrations of the laser cavity mirrors or temperature fluctuations, the linewidth is always a finite amount, e.g., a Gaussian shape. This means that even if the pulses are overlapped, at each moment the frequencies (or wavelengths) may be different, i.e., each pulse envelop contains a finite amount of frequencies. This will affect the interference visibility, which in turn will increase QBER and limit the key rate of TF-QKD. 

\subsubsection{Laser locking techniques}

To realize TF-QKD, crucial requirements on Alice and Bob's sources is that they are both monochromatic, i.e., very thin linewidth and share near identical wavelength value. The corresponding phase drift should be extremely steady, e.g., as shown in Fig. \ref{fig:beating}. In classical optics, the technology for achieving phase stability and narrow laser spectral width is very mature, where lasers with linewidths in the order of mHz are already available \cite{Zhang:19tfqkd}. Here, we can adopt similar techniques, where we lock Alice and Bob's lasers to a standard source, which can be laser source with very thin linewidth, an ultra-stable high-finesse cavity, or an optical clock, and tune the laser wavelengths to a near identical value and even thinner linewidths. 

\begin{figure}[h!]
\centering
\includegraphics[width=12cm,trim={0.8cm 0.8cm 0.8cm 1.1cm},clip]{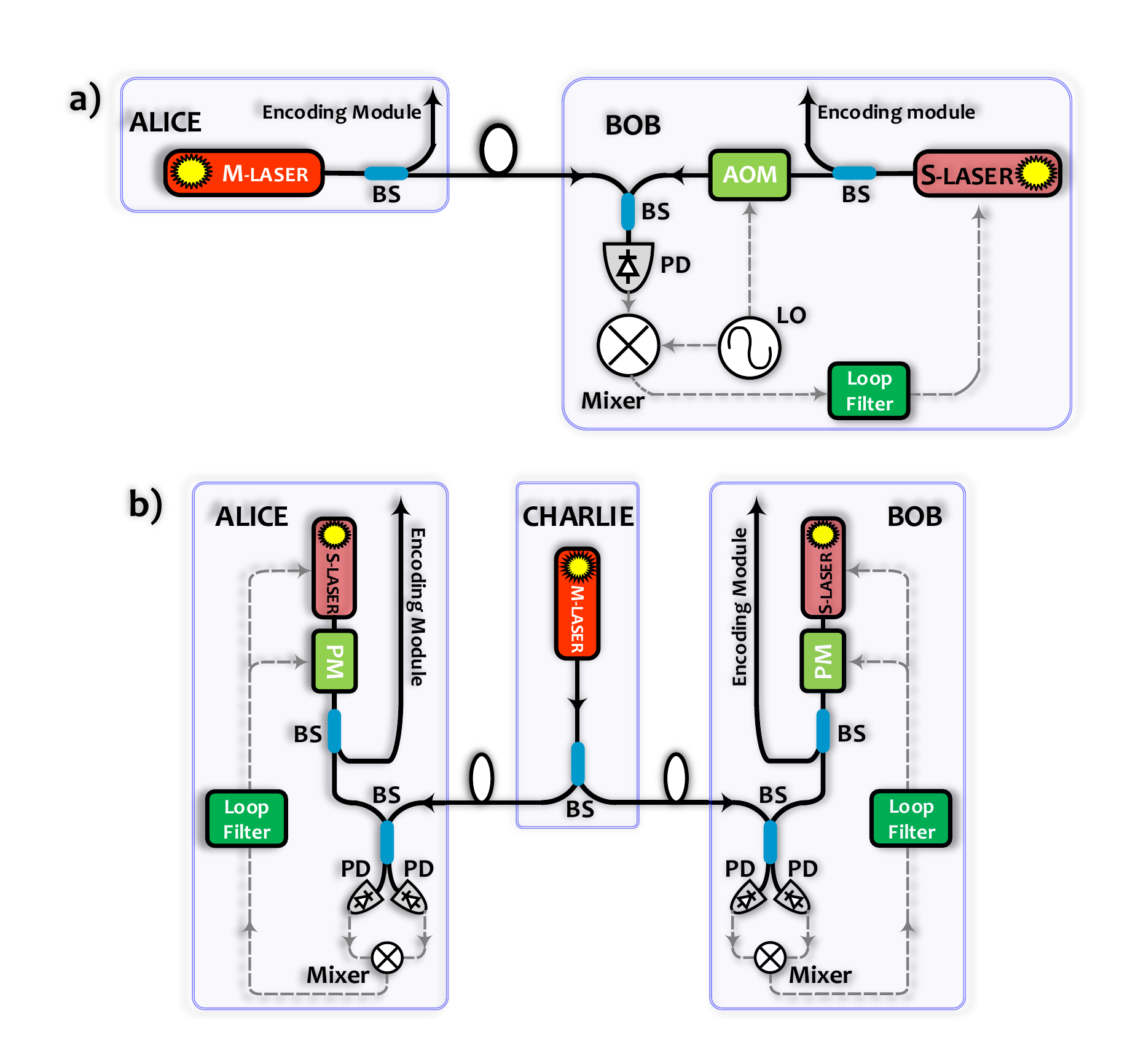}
\caption{Schematics for locking Alice and Bob's lasers using OPLL. a) Locking Bob's laser to Alice's \cite{Minder2019}. b) Locking Alice and Bob's laser to Charlie's \cite{wang19}. M-Laser: master laser; S-Laser: slave laser; AOM: acousto-optic modulator; PD: photodiode; LO: local oscillator; BS: beam splitter. The black solid lines represent fibers, and the dashed gray lines represent electrical signals.}
\label{fig2:OPLL}
\end{figure}

Thus far, in nearly all demonstrations of TF-QKD were realized by locking Alice and Bob's lasers onto a third source \cite{wang19,Fang2020}, or locking Bob's source to Alice's \cite{Minder2019,liu19,chen509snstfqkd}, which are CW lasers with linewidths in the kHz or sub-kHz level. A general technique is through using optical phase locking loop techniques (OPLL) \cite{Ferrero:08}, as shown in Fig. \ref{fig2:OPLL}. Specifically, the lasers to be locked are to interfere with a reference laser beam, the former being the slave laser and the latter being the master laser. The beating signals from PDs together with a mixer form a phase detector, which returns the electrical signal proportional to the phase error. This electrical signal will then be filtered and sent as a feedback signal to the slave laser to complete the homodyne loop operation, as shown in Fig. \ref{fig2:OPLL}b. When an LO signal that is  fixed at a frequency offset is added, the beating signal can be compared with that from the phase detector, and the operation becomes the heterodyne phase-lock loop. 
Phase modulators may be added to further lower the phase noise. When using this technique to lock Alice and Bob's lasers, which have linewidth of $2$ kHz, to a reference beam of $0.1$ kHz linewidth, the visibility of a direct local interference (no fibers between Alice to Charlie and Bob to Charlie) can be as high as $99.78\%$, which represents near zero-offset frequency between the two lasers \cite{wang19}. 

\begin{figure}[h!]
\centering
\includegraphics[width=14cm,trim={0.8cm 0.8cm 0.8cm 1cm},clip]{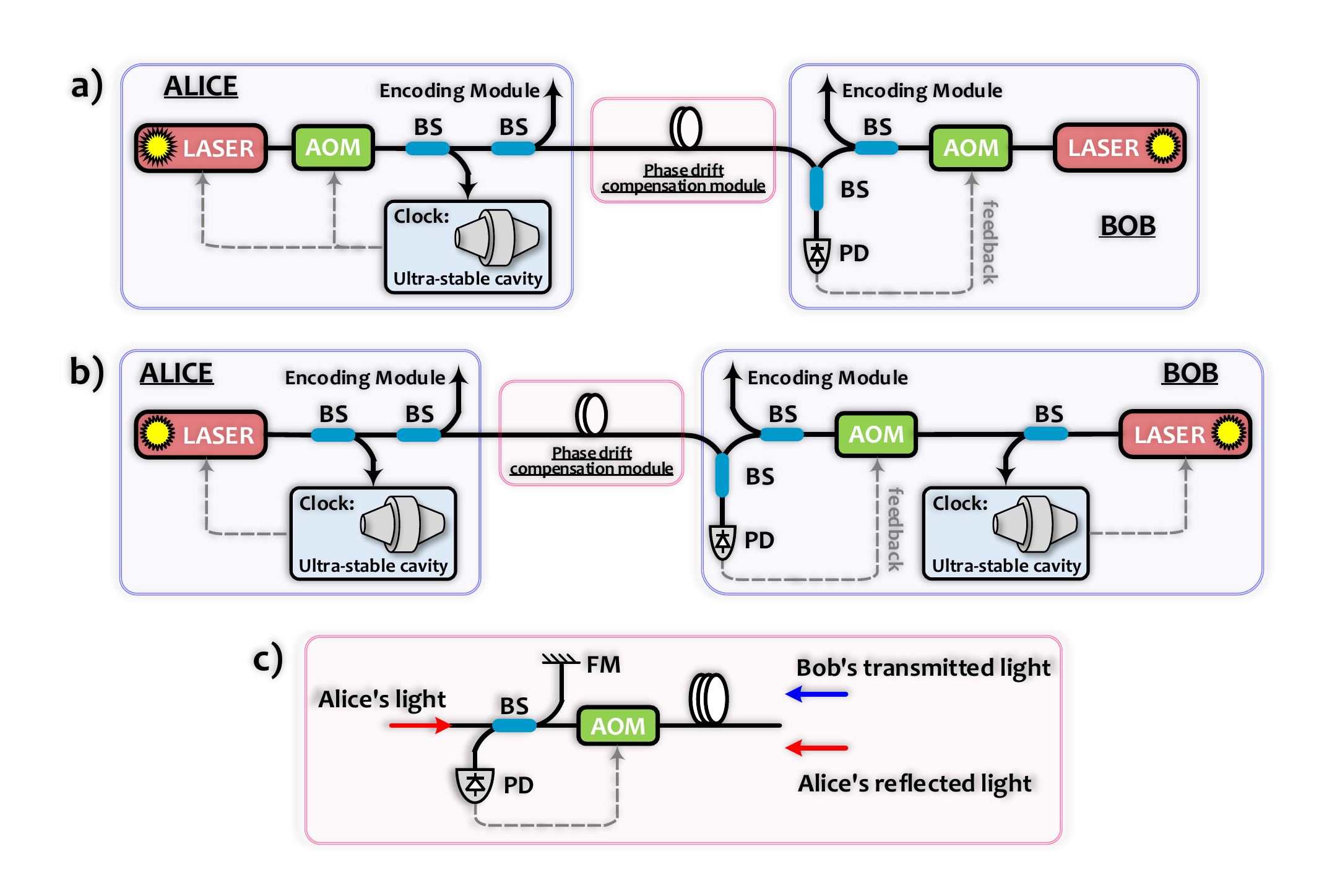}
\caption{Schematics of the time-frequency transfer locking method. a) Locking Alice's laser  wavelength to the ultra-stable cavity, then locking Bob's laser wavelength to that of Alice's \cite{liu19}. b) The wavelength of Alice and Bob's lasers are respectively locked to each of their ultra-stable cavities, then mutually locked their each other using time-frequency dissemination techniques \cite{chen509snstfqkd}. c) Phase drift compensation module \cite{liu19,chen509snstfqkd}. Alice's reflected light is used to generate the beating signal at the PD for the locking schematics of a), while Bob's transmitted light is used for that of locking schematics b). AOM: acousto-optic modulator; BS: beam splitter; PD: photo diode; FM: Faraday mirror. The black solid lines represent fibers, and the dashed gray lines represent electrical signals.}
\label{fig3:TFtransf}
\end{figure}

\begin{figure}
     \centering
     \begin{subfigure}[b]{0.49\textwidth}
         \centering
         \includegraphics[width=8.5cm,trim={8.2cm 5cm 9cm 5cm},clip]{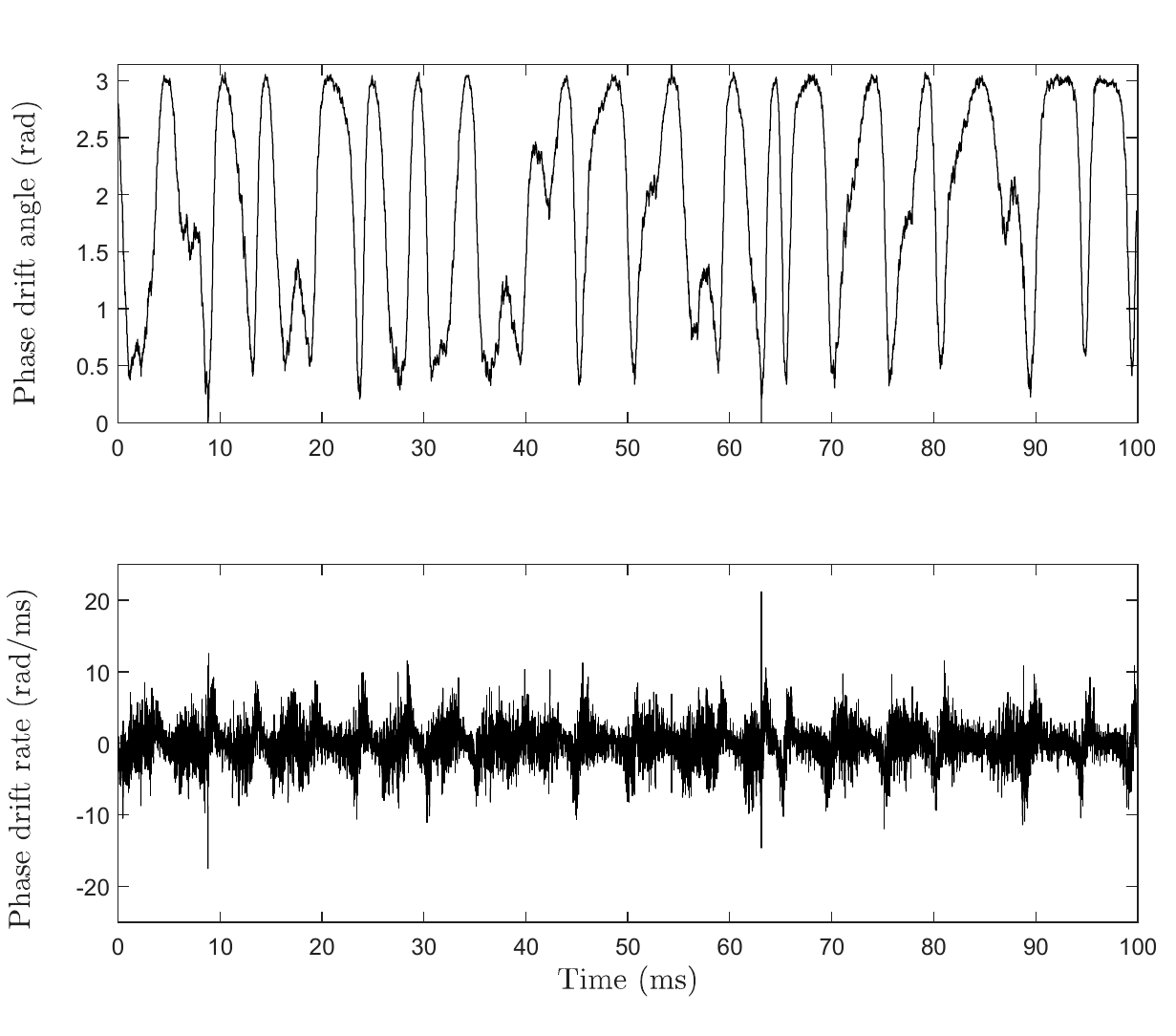}
         \caption{With $0$ km fiber between Alice and Bob, the phase drift rate follows the Gaussian distribution with a standard deviation of $2.72$ rad/ms. }
         \label{fig:test1}
     \end{subfigure}
     \hfill
     \begin{subfigure}[b]{0.49\textwidth}
         \centering
         \includegraphics[width=8.5cm,trim={8.5cm 5cm 9cm 5cm},clip]{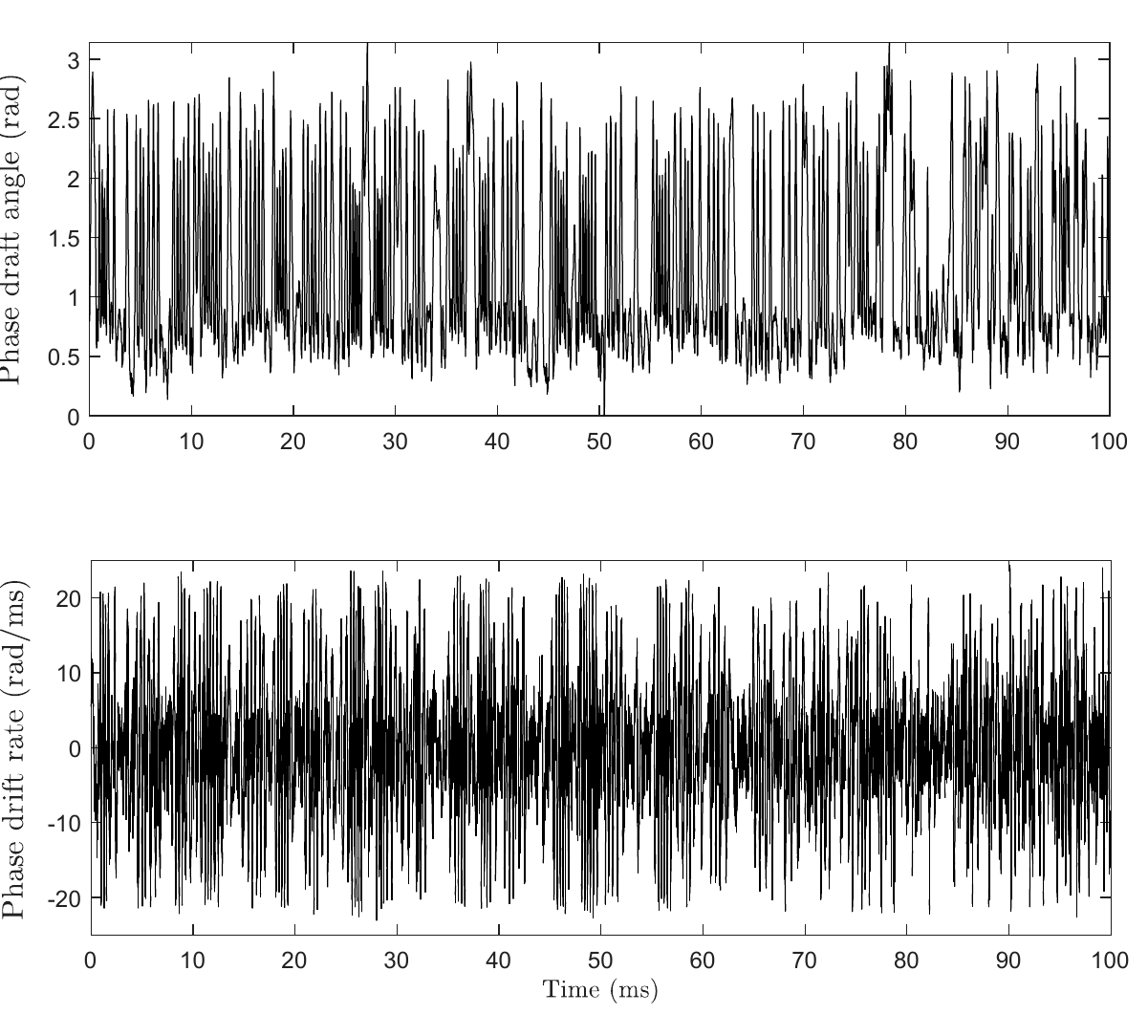}
         \caption{With $509$ km fiber between Alice and Bob, the phase drift rate follows the Gaussian distribution with a standard deviation of $9.58$ rad/ms. }
         \label{fig:test2}
     \end{subfigure}
     \caption{Relative phase drift between Alice and Bob's lasers using the time-frequency transfer method. Reproduced with permission.\textsuperscript{[Ref. \cite{chen509snstfqkd}]} Copyright 2020, APS.}
     \label{fig:trans}
\end{figure}

To increase wavelength stability and decrease laser linewidth for higher coherence, one can apply a fiber-optic time-frequency transfer based method, where Alice's CW laser was used as the seed laser to lock Bob's \cite{liu19}, as shown in Fig. \ref{fig3:TFtransf}a. First, Alice's laser is locked to an ultra-stable cavity to suppress its linewidth to $\sim10$ Hz, using the Pound-Drever-Hall technique \cite{pound1946,Drever1983}. Then, a portion of Alice's beam is sent to Bob via a third fiber, where a local beating (Alice and Bob directly connected) of the locked lasers showed a phase drift rate of $5.8$ rad/ms \cite{liu19}. This method can be optimized to allowing Bob to also have an ultra-stable cavity to lock his laser, as shown in Fig. \ref{fig3:TFtransf}b, which would further improve the phase drift rate of the lasers to $2.72$ rad/ms and laser linewidth of $\sim1$ Hz \cite{chen509snstfqkd}. At $509$ km, the relative phase drift between Alice and Bob's lasers are shown in Fig. \ref{fig:trans}. To simulate field conditions, the length of the fiber directly connecting Alice and Bob will be the same as the total transmission distance, so any phase noise existing in this transmission channel should be compensated. This is done by using the setup shown in Fig. \ref{fig3:TFtransf}c, where by analyzing the beating signals of Alice's local light from the FM and Bob's transmitted light or Alice's reflected light, the phase drifts are compensated using the AOM. 

\begin{figure}
     \centering
     \begin{subfigure}[b]{0.8\linewidth}
         \centering
         \includegraphics[width=14cm,trim={0.4cm 0.4cm 0.8cm 0.8cm},clip]{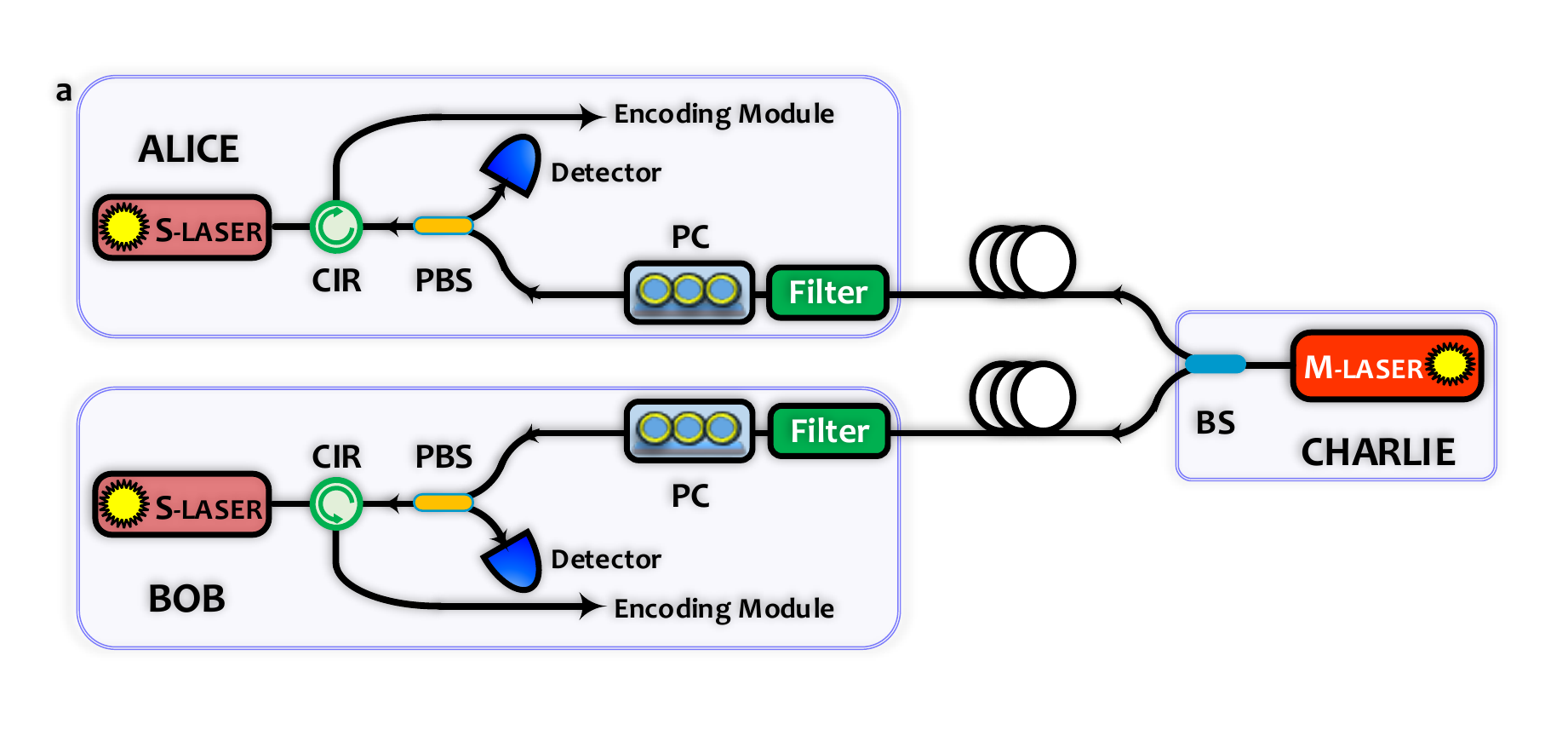}
                  \label{fig:test1}
     \end{subfigure}
         \begin{subfigure}[b]{0.8\linewidth}
         \centering
         \includegraphics[width=12cm,trim={0.8cm 5cm 0.8cm 4.5cm},clip]{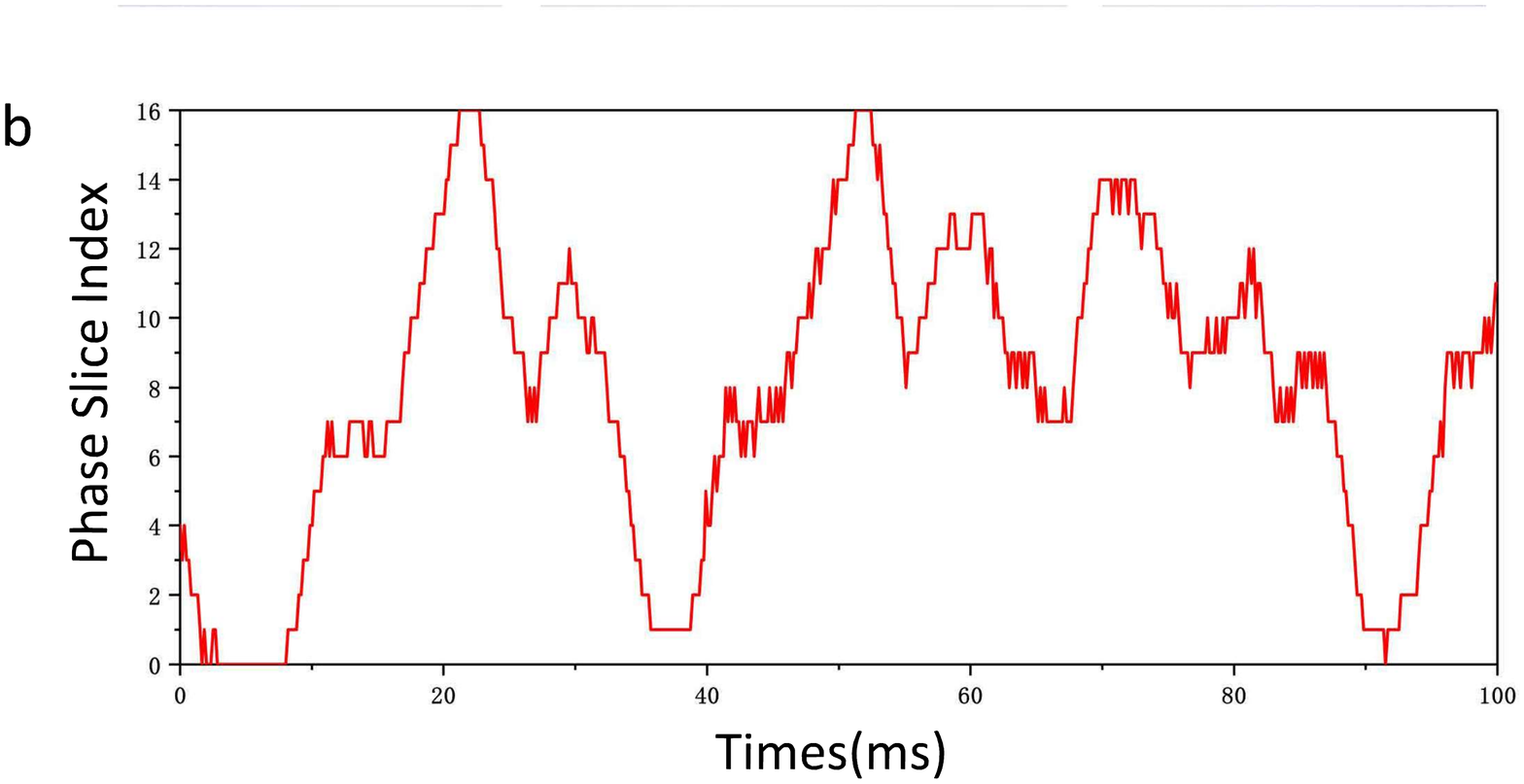}
                  \label{fig:test2}
     \end{subfigure}
     \caption{a) Schematics of the laser injection method. b) Phase drift of Alice and Bob's lasers with $0$ km fiber in between. Reproduced with permission.\textsuperscript{[Ref. \cite{Fang2020}]} Copyright 2020, Nature.}
     \label{fig4:laserinj}
\end{figure}	

Another method of laser injection \cite{liu2020injection,Comandar2016} was applied in \cite{Fang2020}. Here the seed laser is set at Charlie, as shown in Fig. \ref{fig4:laserinj}. By directly injecting the seed laser to Alice and Bob's slave lasers, the phases of both Alice and Bob's lasers are locked to that of Charlie's. Directly interfering Alice and Bob's locked lasers shows that the phase drift rate can be as low as $0.62$ rad/ms. This method was also demonstrated in a realistic scenario, where the master laser signals travel through long fibers to reach the slave lasers, though the locking efficiency was shown to be decreasing as transmission length increases. 

Through the above demonstrations, one can see that that independent lasers of Alice and Bob can be indeed locked to show ``twin'' characteristics. From a security perspective, these three methods can be categorized into two types. With the first two methods, although an extra fiber channel is added between Alice and Bob, the pulses used for key generation are still solely prepared by Alice and Bob, so Eve is only in control of state measurement. Even when a third laser source is introduced at Charlie, this light is only for comparing the wavelengths between Alice and Bob. With the third method, however, the master laser is the incident light for Alice and Bob's lasers. If a malicious Eve has control of this light, she can theoretically tamper it by encoding some information before sending it to Alice and Bob. After Alice and Bob encode their phase information, Eve may use her initial encoding to learn some key information from the pulses. This may be dealt with by monitoring the master laser light in case of tampering, though the exact form of loopholes that can be exploited by Eve and the extent of such attacks on the security of experimental TF-QKD is still an open question.

\subsection{Fiber induced phase noise }

The optical phase of the light field in very sensitive to the fiber environment, such as fiber temperature, stress, vibrations, and length. Generally, temperature affects phase the most, as ordinary single mode fibers (SMF) have a positive thermal expansion coefficient that causes transmission time delay to vary in proportion to temperature when they are used for signal transmission. Such environmental factors cannot be controlled freely, especially over long distances, and the resulted phase noise tends to corrupt the wave front of the original light field and cause the original delta-type Hz-level optical spectrum to broaden, which will in turn disrupt the steady interference between two distant sources. At $200$ km of fiber, the phase drifting was measured to be $2.8$ rad/ms, and by $400$ km, this value nearly doubles to $5.5$ rad/ms. The drifting rate does not increase linearly, as for $600$ km, the rate is $7.1$ rad/ms, and by $800$ km, the drifting can be as high as $15.7$ rad/ms \cite{Fang2020}.

\begin{figure}[h!]
\centering
\includegraphics[width=14cm,trim={1cm 1.7cm 1cm 1.3cm},clip]{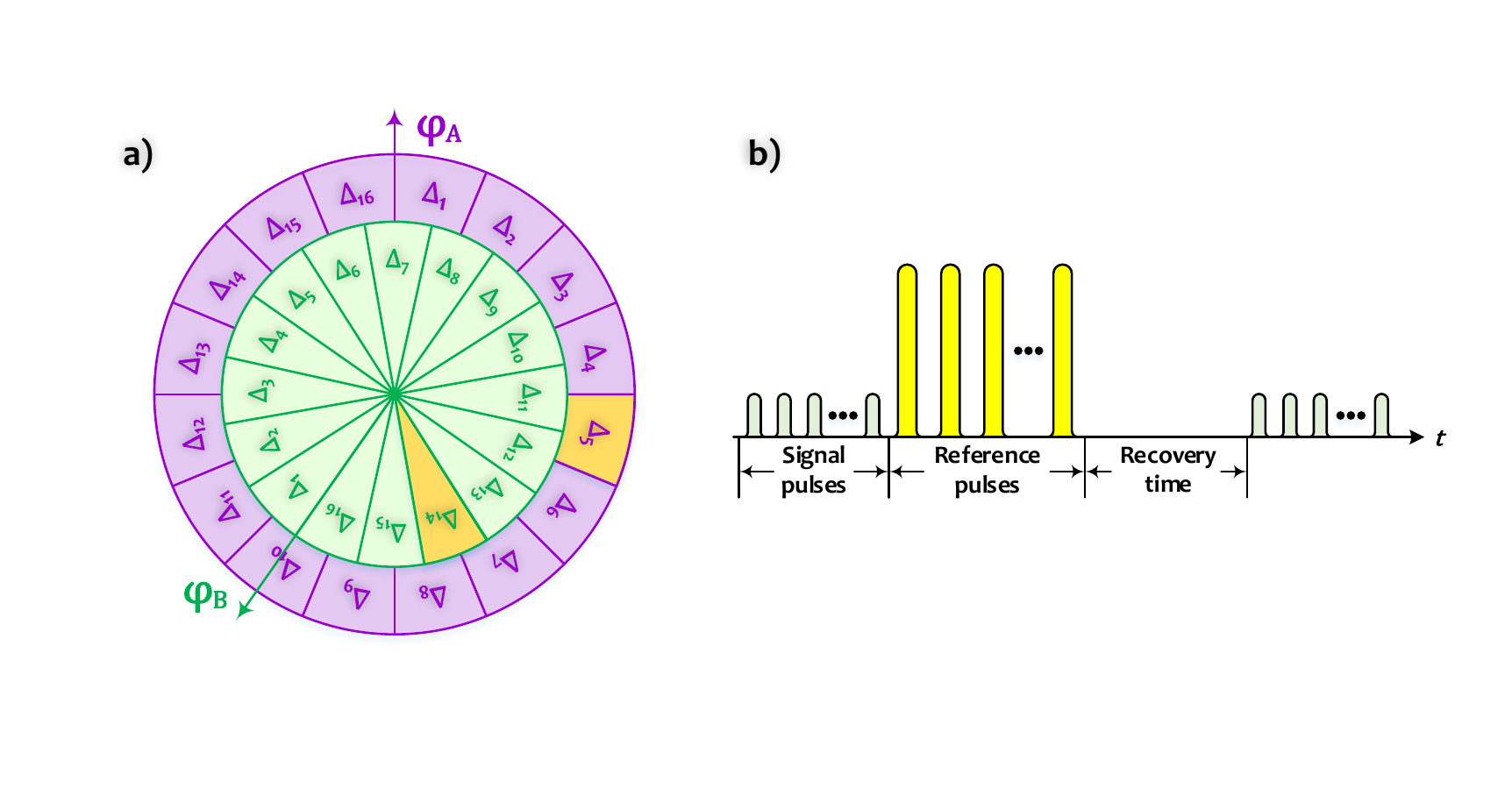}
\caption{a) Phase slices for PM-QKD, where the number of slices $M$ is set to $16$ as an example. After transmission, Alice and Bob's phase references differ by $\Delta\varphi$. b) Pulse train sequence for compensating fiber-induced phase noise.}
\label{fig5:phaseslice}
\end{figure}

For a typical TF-QKD or PM-QKD protocol, Alice and Bob each encode a random phase $\rho_{A}$ ($\rho_{B}$) and a bit phase $\alpha_{A}$ ($\alpha_{B}$) to their respective pulses (suppose the sources are perfectly phase-locked). After transmission, the pulses would also have obtained a phase shift 
from the fiber $\tau_{A}$ ($\tau_{B}$), so the relative phase for each pulse would be $\Delta\varphi=\left(\rho_{A}+\alpha_{A}+\tau_{A}\right)-\left(\rho_{B}+\alpha_{B}+\tau_{B}\right)$. Thus, when Alice and Bob's beams interfere at the BS, the intensities of single photon detectors $D_0$ and $D_1$ are respectively $I_{0}=(1+\cos\Delta\varphi)/2$ and $I_{1}=(1-\cos\Delta\varphi)/2$ after normalization. This indicates that the interference results detected are actually resulted from the combined effects of phase encoding and fiber induced phase noise. For simplicity, consider the case where $\rho_{A}=\rho_{B}$ and $\alpha_{A}=\alpha_{B}$, then $\Delta\varphi=\tau_{A}-\tau_{B}$, as shown in Fig. \ref{fig5:phaseslice}a. The interference visibility can be defined by $V=\frac{I_{0}-I_{1}}{I_{0}+I_{1}}$, then a estimate of the induced error can be written as $\epsilon=\left(1-V\right)/2=\frac{I_{1}}{I_{0}+I_{1}}$. For a fiber channel of up to $550$ km, the phase drifting may reach $6.0$ rad/ms \cite{Lucamarini2018}, which according to the above estimation would cause $\sim2\%$ extra error rate and severely affect the key rate. 

In conventional optical communications, the issue of phase noise induced by transmission fiber may be dealt with by using induced phase noise cancellation methods, such as first measuring the noise using a single thin linewidth laser beam and performing a double-pass heterodyne measurement, then incorporate the results into the affected data \cite{ma1995spie}. Unfortunately, analysis time for this phase noise cancellation method is proportional to the length of the tested fiber, so as the fiber lengthens, or other environmental factors randomly change, this method may not be able to keep up with the fast variation of the fiber. Nevertheless, we can analyze the measured interference results and see that within a short period of time, usually several $\mathrm{\mu}$s, the phase drift from the fiber can remain stable. Thus, we can send extra pulses along with signal pulses to monitor the phase drift of fibers for a period of time, as shown in Fig. \ref{fig5:phaseslice}b. Here, the signal pulses are dim pulses for key generation and error estimation, while the reference pulses are bright pulses for analyzing the phase noise of the transmission fibers. Also, after the strong reference pulses, there is a recovery time period for the single photon detectors (SPD) where no pulses are sent. To combat the fiber phase drift, one can use two methods depending on the TF-QKD protocol implemented.

\subsubsection{No phase post-selection }

\begin{figure}[h!]
\centering
\includegraphics[width=12cm,trim={0.5cm 0.5cm 0.5cm 0.5cm},clip]{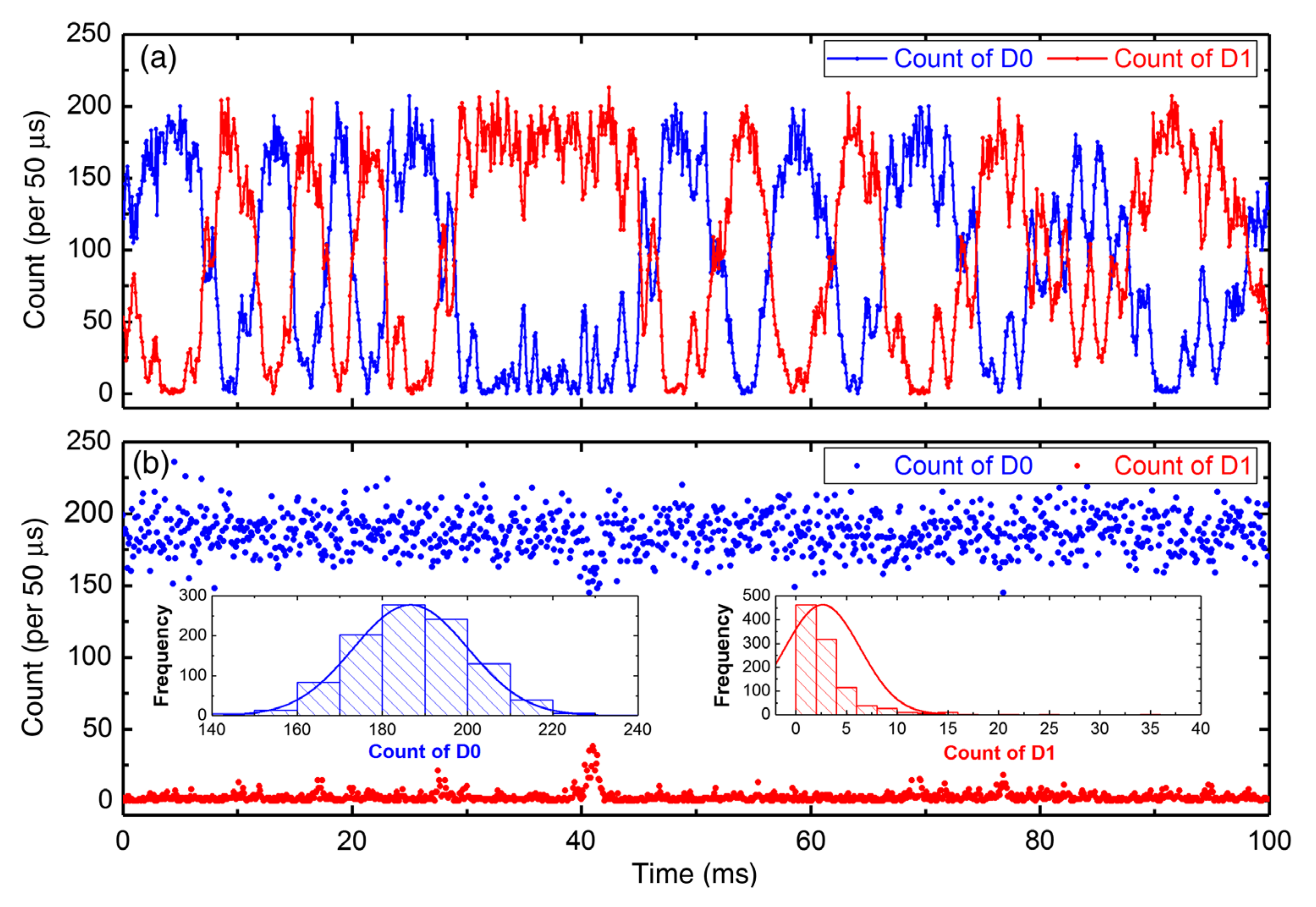}
\caption{Interference results of reference pulses due to fiber induced phase drift at $300$ km. a) Without PM; b) with PM. Reproduced with permission.\textsuperscript{[Ref. \cite{wang19}]} Copyright 2019, APS. }
\label{fig:300kmphase}
\end{figure}

For experiments implementing the no phase post-selection (NPP) TF-QKD, the phase references of Alice to Charlie and Bob to Charlie must be calibrated before the protocol \cite{cui19,Curty2019}. In Ref. \cite{wang19}, for example, Alice and Bob each send $50$ $\mu$s of weak quantum signals followed by $50$ $\mu$s of strong optical pulses, which are unmodulated and to interfere at Charlie. A phase modulator is placed before each of Charlie's detectors, where their voltages are adjusted to maximized the interference visibility in real-time. With this method, as shown in Fig. \ref{fig:300kmphase}, the interference visibility can be effectively increased,  and at $300$ km the visibility is observed $\sim97.21$\%. 

The advantage of using this method is obvious, for it not only is simple to realize, but also eliminates complicated phase post-selection process and uses all data for key generation. However, as the fiber lengthens, fiber phase drift may be too fast to recover with the current active phase compensation technique. To improve the phase feedback, the reference pulse intensities should be increased, and the duration of bright and dim pulses should be shortened.

\subsubsection{With phase post-selection }

For protocols such as PM-QKD and SNS-TF-QKD, the phase noise from fibers are not compensated, but instead measured using interference results of bright pulses and incorporated into relevant calculations. First, one estimates the phase noise. Since single photon detectors are used, the intensities of $I_{0}$ and $I_{1}$ can be observed calculating the probability of detector clicks in each phase shift estimation round. By building error models, one can also monitor the success rate using this phase noise estimation method. Then, one can take the estimated phase back into $\Delta\varphi$, and cancel the effects from fiber to compare whether the global random phases of Alice and Bob are in the same slice and calculate the correct value of QBER \cite{liu19}. 

Depending on the locking method for Alice and Bob's lasers, the settings for pulses trains are different, namely as in the duration of each pulse train, the duty cycles of each type of pulse, and the reference pulse phase modulation and intensity modulation. This post-processing method can in principle analyze fibers under any circumstance, yet just as the issue with NPP-TF-QKD, longer transmission distances would require the high accuracy estimations, which would require more reference pulses with higher optical power, and may in turn reduce the duty cycle of quantum signals, cause more scattering noise photons in the channel, and affect the QBER.

\subsubsection{Plug-and-play TF-QKD }

\begin{figure}[h!]
\centering
\includegraphics[width=8cm,trim={0.5cm 0.8cm 0.5cm 0.8cm},clip]{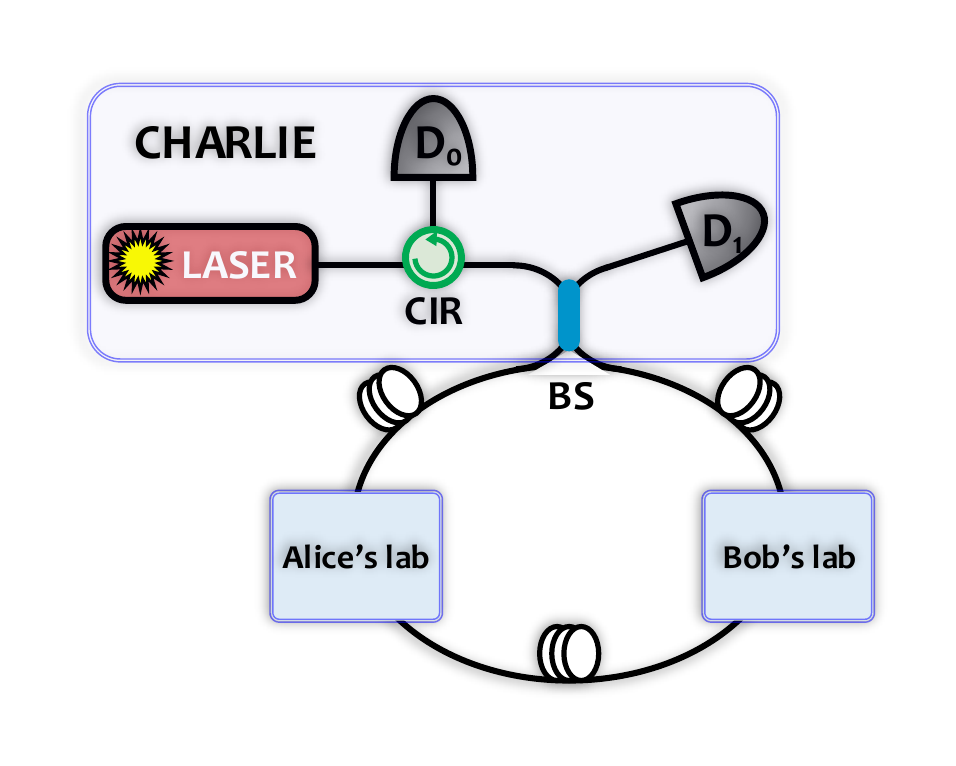}
\caption{Schematics for NPP-TF-QKD using a Sagnac interferometer \cite{zhong19}. CIR: circulator; BS: beam splitter; $D_0$: detector 0; $D_1$: detector 1.}
\label{fig6:sagnac}
\end{figure}

It was proposed in Ref. \cite{zhong19} that a plug-and-play system, combined with a Sagnac-like interferometer schematic, as shown in Fig. \ref{fig6:sagnac}, can be used to avoid the issues of laser locking and fiber induced phase noise estimations to realize NPP-TF-QKD. The idea was that the common-path nature of the Sagnac interferometer automatically compensates for phase fluctuations of the two fields from Alice and Bob, without requiring the phase feedback to align the phase references. Also, by using the idea of the ``plug-and-play'' QKD system \cite{plugnplay}, independent lasers are no longer required for Alice and Bob, hence also phase locking techniques are eliminated. In this sense, this experiment is quite unique, as all others implement the original TF-QKD schematics where Alice and Bob each has an independent laser to interfere at Charlie. This scheme was further studied under asymmetric channel settings \cite{zhong2020proof}. 

However, the plug-and-play schematics also has its limitations. The most serious issue is that the only source in this schematic is placed at Charlie, which may be controlled by Eve. This can open security loopholes similar to those with the laser injection method, as the quantum states are generated and measured by Eve, so further security analysis of the plug-and-play setup is necessary. Experimentally speaking, the automatic phase compensating for Sagnac loops is conditioned on the fact that the phase of the loop remains relatively stable. In this work, a $17$ km loop demonstrated stable interference with visibility $>99\%$, but as the fiber lengthens, active phase compensation may still be required to deal with the fast phase drift. 

\subsection{Other considerations }

\subsubsection{Increasing maximum count rate for single photon detectors }

In TF-QKD, as the bright reference pulses are time-multiplexed with the quantum signal pulses, SPDs detect both signals together. However, for superconducting nanowire (SN) SPDs, which are preferred in long distance QKD experiments due to their high detection efficiencies, the kinetic inductance, which is associated with the maximum count rate of the detector, is inversely proportional to its detection efficiency \cite{kerman2006}. In some cases, an SNSPD with $65\%$ efficiency can only detect photons of up to $\sim2$ MHz before the nanowire stops responding linearly to incoming photons \cite{lv2018snspd}. This may create problems in TF-QKD, as bright reference pulses are required for accurate phase estimations, which may be beyond the maximum count rate for the SNSPDs and cause the nanowire to have the latching effect \cite{annunziata2010}. New developments in nanowire configuration is necessary to reduce the kinetic inductance without compensating detection efficiency. Other approaches can be used such as inserting a $50$ $\mathrm{\Omega}$ shunt resistor between the DC arm of the bias tee and the ground to prevent the detector latching at high photon count rates \cite{Liu_2012}. 

\subsubsection{Noise }

For some TF-QKD protocols, e.g., the sending-or-not-sending protocol, strict requirements are placed on the SPD dark counts. To reach distances up to $500$ km, SNSPDs with dark counts as low as $<10$ counts per second have been developed \cite{chen509snstfqkd}. 

In practice, other factors can also induce noise to the QKD system. This is mainly caused by the strong reference pulses sent from Alice and Bob to Charlie for phase estimation. For a typical TF-QKD type experiment, the receiving photon count rate for the strong reference pulses at each detector is $1\sim2$ Mcps \cite{Fang2020}, which is much higher than the detected count rate for quantum signals, so several nonlinear optical effects may occur, such as Rayleigh scattering, spontaneous Brillouin scattering, and spontaneous Raman scattering \cite{agrawal2007nonlinear}, and noise photons will be introduced. 
Here, Rayleigh scattering is elastic scattering, where the wavelength is same as the propagating pulses. Due to many impurity defects along the optical fiber, light can be backward scattered with Rayleigh scattering in one point in the fiber, and then backward scattered by Rayleigh scattering again at a different impurity, which is called Re-Rayleigh scattering in optical fiber. For reference pulses of $\sim2$ Mcps receiving rate, after 500 km (250 km for Alice to Charlie and Bob to Charlie), the noise photon rate is in the order of $10^{-8}$ \cite{chen509snstfqkd}. This is inevitable with the use of optical fibers, and along with reflection of fiber facets in single photon detectors, will severely limit the performance of long distance TF-QKD. Such a challenge may be avoided by replacing Alice and Bob's single frequency lasers with other sources such as the optical frequency comb \cite{Fortier2019}.

Meanwhile, spontaneous Brillouin scattering and spontaneous Raman scattering are forward inelastic scattering, which compared to Rayleigh scattering, the intensity of the former is $2\sim 3$ orders of magnitude lower with a frequency shift less than 1 GHz, while the intensity of the latter is $3\sim 5$ orders of magnitude lower with a THz frequency shift \cite{agrawal2007nonlinear}. Brillouin scattering can be eliminated using time-division-multiplexing and setting longer recovery time in the pulse sequence to separate the quantum and reference signals, while Raman scattering can be filtered using narrow pass-band filters \cite{Mao2018QKD}.


\subsection{Results }

Applying the above state-of-the-art techniques, experimental demonstrations have confirmed that the PLOB bound can be overcome with TF-type QKD protocols, with \cite{wang19,liu19,Fang2020,chen509snstfqkd} and without \cite{Minder2019,zhong19,zhong2020proof} actual fibers. To break the linear bound, the shortest distance is obtained at $300$ km, where the key rate is $6.46\times10^{-6}$ and $1.96\times10^{-6}$ without \cite{wang19} and with finite size effects considered \cite{liu19}, respectively.  Interference is proven stable, where visibility can remain above $\sim96\%$ for over $1000$ s \cite{wang19}. Thus far, laboratory based experiments of TF-QKD type protocols have reached $502$ km using ultra-low loss fiber, with key rate of $8.43\times10^{-10}$ \cite{Fang2020}. To further break the absolute linear bound, where no device imperfections are considered, key rate of $6.19\times10^{-9}$ can be obtained over $509$ km transmission \cite{chen509snstfqkd}. These are the longest QKD demonstrations for fiber-based experiments without repeaters. We summarize the main experimental results in Table \ref{tab1}.

\begin{table}
\caption{List of experiments overcoming the linear secret key capacity bound. Att: attenuation.}
\centering
\label{tab1}
\begin{tabular}{ccccccc}
\hline 
\hline 
\textbf{Reference} & \textbf{Protocol} & \textbf{Clock rate} & \textbf{Fiber Length} & \textbf{Key rate} & \textbf{Finite size} & \textbf{Note} \tabularnewline

\hline 
Minder, \textit{et al.}, 2019 \cite{Minder2019} & TF-QKD & $2$ GHz & -  & $2.25\times10^{-8}$ & No & $90.8$ dB Att\tabularnewline

Wang, \textit{et al.}, 2019 \cite{wang19} & NPP-QKD & $1$ GHz & $300$ km & $6.46\times10^{-6}$ & No &\tabularnewline

Liu, \textit{et al.}, 2019 \cite{liu19} & SNS-QKD & $33.3$ MHz & $300$ km & $1.96\times10^{-6}$ & Yes &\tabularnewline

Zhong, \textit{et al.}, 2019 \cite{zhong19} & NPP-QKD & $10$ MHz & -  & $1.75\times10^{-5}$ & No & $55.1$ dB Att \tabularnewline

Fang, \textit{et al.}, 2020 \cite{Fang2020} & PM-QKD & $312.5$ MHz & $502$ km & $8.43\times10^{-10}$ & Yes &\tabularnewline

Chen, \textit{et al.}, 2020 \cite{chen509snstfqkd} & SNS-QKD & $33.3$ MHz & $509$ km & $6.19\times10^{-9}$ & Yes &\tabularnewline
 
Zhong, \textit{et al.}, 2020 \cite{zhong2020proof} & NPP-QKD & $10$ MHz & - & $3.17\times10^{-7}$ & Yes & $56$ dB Att \tabularnewline
\hline 
\end{tabular} 
\end{table}

\section{Conclusions and outlook}

In summary, we have surveyed the theoretical and experimental developments in the past few years on overcoming the linear secret key capacity bound. By applying the TF-QKD protocol, one can beat the PLOB bound and achieve higher key rates and longer transmission distances than existing single-photon based QKD protocols. The unconditional security of TF-QKD has been proven, and experiments have further verified that TF-QKD can beat the PLOB and reach distances of up to $500$ km, which gives promise to practical applications. 

Due to its obvious advantages in high key rate and long transmission distance, TF-QKD is highly anticipated to play more prominent roles in practical applications. However, quite a few challenges are yet to be overcome. 
One important factor to be considered is the security of the optical source in experimental setups. For instance, the security of Charlie’s laser in the laser injection method and the Sagnac loop setup should be re-examined before further applications, and if loopholes do in fact exist, the question of how to patch them should be answered. Also, although TF-QKD is immune to side-channel attacks, possible hacking strategies towards the source should be considered \cite{Pereira2019}. 

Another factor to consider for practical applications of TF-QKD is that currently experiments are all performed in laboratories, where the experimental conditions differ greatly from actual field settings. The most common difference is that the fiber channels from Alice to Charlie and Bob to Charlie in real-life are likely to be asymmetric. Currently, asymmetric TF-QKD has only been studied with the Sagnac loop setup \cite{zhong2020proof}. With other setups, technically the issue can be solved by adding fiber spools to compensate the fiber difference, or applying TF-QKD protocols optimized for asymmetric settings. However, it is still vital to study how the asymmetry of transmission channels will affect the interference visibility and error rate. 

Also, for the sake of demonstration, essential components such as synchronization were mostly simplified in reported TF-QKD experiments, e.g., in Ref. \cite{Fang2020} synchronization was achieved with electrical signals. In an actual field application, the most common way to synchronization all systems of Alice, Bob, and Charlie, is through optical synchronization \cite{Mao2018QKD}. However, this would require sending an extra optical signal of a different wavelength from quantum signals along the quantum channel using wavelength-division multiplexing methods. Here, nonlinear optical noise photons may also be generated based on how wide the wavelengths differ and how much the signals’ optical powers differ, so further detailed investigations are required to evaluate the wavelength, repetition rate, and emission power of the optical synchronization signal.

Meanwhile, currently the longest transmission for TF-QKD is $\sim500$ km. To achieve even longer distances, on the one hand a critical factor is the use of ultra-thin linewidth lasers at Alice and Bob's sites. For a laser of center wavelength at 1550 nm and 200 Hz linewidth, its coherence length is $\sim500$ km, so even thinner linewidths are required for longer transmissions. Currently, only the time-frequency transfer method can achieve such a thin linewidth \cite{chen509snstfqkd}, but the experimental setup is extremely complex. Other simpler techniques such as laser injection and OPLL should be optimized to achieve thinner linewidths. On the other hand, low signal-to-noise ratio, which is due to attenuation of QKD signal pulses, the generation of nonlinear optical noise photons, and the dark counts of the detectors, also limits long distance TF-QKD. To increase the signal-to-noise ratio, higher system repetition rates are required, which places high demands on the components used in quantum state preparation, i.e., the electrical signals controlling electro-optic modulators. For nonlinear optical noise photons, they are caused by fundamental impurities and material of the optical fiber, so to reduce these effects, new types of fibers should to developed and investigated. Also, ultra-low dark count single photon detectors should be developed to increase the signal-to-noise ratio.

Furthermore, in realistic scenarios TF-QKD may need to co-exist with classical optical communication signal along a single fiber \cite{Mao2018QKD}. This would further lower the signal-to-noise ratio, as classical communication signals are much higher in optical power and would generate scattering noise photons within the wavelength bandwidth of the quantum signal. The amount of noise photons is related to the type of fiber, the emission power of the classical communication signals, and the relative direction of propagation between QKD and classical signals. Such an implementation requires further investigation. 

Another important application is networking. Similar to MDI-QKD, TF-QKD can be naturally extended to a star-type network implementation where Charlie is the untrusted central node \cite{tang2016measurement}. This will bring more challenges on top of a field trial, where more stabilization systems may be required and laser locking will be necessary for multiple users.

Finally, many interesting topics are still waiting to be covered both theoretically and experimentally. For example, the key rate of TF-QKD is still a long way from the single repeater bound, as shown in Fig. \ref{fig0:keyrate}, so it is interesting to investigate how the key rate scaling can be increased even more from a protocol perspective. Meanwhile, a continuous-variable (CV) version of TF-QKD is absent, and given the advantages of CV-QKD \cite{pirandola2019advances}, it may open a new potential area of research for QKD overcoming the linear key capacity bound. In addition, studies in simplifying current TF-QKD systems, particularly laser-locking modules, chip-based implementation \cite{sibson2017chip,wei2019high}, and applying TF-QKD to other quantum communication protocols \cite{zhao2020} can be anticipated to promote practical applications of TF-QKD.

\medskip

\medskip
\textbf{Acknowledgements} \par 

We thank Profs. X. Ma, F. Xu, and J. Majer, and H.-Y. Zhou, J.-P. Chen, X.-T. Fang, C. Jiang, and H. Liu for enlightening discussions, as well as the anonymous reviewers for their constructive comments and suggestions in improving this review. 
Y. M. especially thank Y.-Z. Zhen for numerous fruitful discussions, encouragements, and support over many late night revisions throughout the entire project. 

This work is supported by the National Key R \& D Program of China (No. 2017YFA0303903), and the National Natural Science Foundation of China (Grant No. 61875182).

\medskip

%
\bibliographystyle{MSP}
\bibliography{sample}

\begin{thebibliography}{10}
\providecommand{\url}[1]{\texttt{#1}}
\providecommand{\urlprefix}{URL }

\bibitem{Scarani2009RMP}
V.~Scarani, H.~Bechmann-Pasquinucci, N.~J. Cerf, M.~Du\ifmmode~\check{s}\else
  \v{s}\fi{}ek, N.~L\"utkenhaus, M.~Peev,
\newblock \emph{Rev. Mod. Phys.} \textbf{2009}, \emph{81} 1301.

\bibitem{BB84}
C.~H. Bennett, G.~Brassard,
\newblock In \emph{Proceedings of IEEE International Conference on Computers,
  Systems, and Signal Processing}. IEEE, New York, \textbf{1984} 175--179.

\bibitem{xu2019secure}
F.~Xu, X.~Ma, Q.~Zhang, H.-K. Lo, J.-W. Pan,
\newblock \emph{Rev. Mod. Phys.} \textbf{2020}, \emph{92} 025002.

\bibitem{pirandola2019advances}
S.~Pirandola, U.~Andersen, L.~Banchi, M.~Berta, D.~Bunandar, R.~Colbeck,
  D.~Englund, T.~Gehring, C.~Lupo, C.~Ottaviani, et~al.,
\newblock \emph{arXiv preprint arXiv:1906.01645} \textbf{2019}.

\bibitem{boaron18}
A.~Boaron, G.~Boso, D.~Rusca, C.~Vulliez, C.~Autebert, M.~Caloz, M.~Perrenoud,
  G.~Gras, F.~Bussi\`eres, M.-J. Li, D.~Nolan, A.~Martin, H.~Zbinden,
\newblock \emph{Phys. Rev. Lett.} \textbf{2018}, \emph{121} 190502.

\bibitem{yin2016mdi404}
H.-L. Yin, T.-Y. Chen, Z.-W. Yu, H.~Liu, L.-X. You, Y.-H. Zhou, S.-J. Chen,
  Y.~Mao, M.-Q. Huang, W.-J. Zhang, H.~Chen, M.~J. Li, D.~Nolan, F.~Zhou,
  X.~Jiang, Z.~Wang, Q.~Zhang, X.-B. Wang, J.-W. Pan,
\newblock \emph{Physical Review Letters} \textbf{2016}, \emph{117}, 19 190501.

\bibitem{braunstein12}
S.~L. Braunstein, S.~Pirandola,
\newblock \emph{Phys. Rev. Lett.} \textbf{2012}, \emph{108} 130502.

\bibitem{mdiqkd}
H.-K. Lo, M.~Curty, B.~Qi,
\newblock \emph{Phys. Rev. Lett.} \textbf{2012}, \emph{108} 130503.

\bibitem{chen2010metropolitan}
T.-Y. Chen, J.~Wang, H.~Liang, W.-Y. Liu, Y.~Liu, X.~Jiang, Y.~Wang, X.~Wan,
  W.-Q. Cai, L.~Ju, L.-K. Chen, L.-J. Wang, Y.~Gao, K.~Chen, C.-Z. Peng, Z.-B.
  Chen, J.-W. Pan,
\newblock \emph{Optics Express} \textbf{2010}, \emph{18}, 26 27217.

\bibitem{Liao2017nature}
S.-K. Liao, W.-Q. Cai, W.-Y. Liu, L.~Zhang, Y.~Li, J.-G. Ren, J.~Yin, Q.~Shen,
  Y.~Cao, Z.-P. Li, F.-Z. Li, X.-W. Chen, L.-H. Sun, J.-J. Jia, J.-C. Wu, X.-J.
  Jiang, J.-F. Wang, Y.-M. Huang, Q.~Wang, Y.-L. Zhou, L.~Deng, T.~Xi, L.~Ma,
  T.~Hu, Q.~Zhang, Y.-A. Chen, N.-L. Liu, X.-B. Wang, Z.-C. Zhu, C.-Y. Lu,
  R.~Shu, C.-Z. Peng, J.-Y. Wang, J.-W. Pan,
\newblock \emph{Nature} \textbf{2017}, \emph{549} 43.

\bibitem{yuan201810}
Z.~Yuan, A.~Plews, R.~Takahashi, K.~Doi, W.~Tam, A.~Sharpe, A.~Dixon,
  E.~Lavelle, J.~Dynes, A.~Murakami, et~al.,
\newblock \emph{Journal of Lightwave Technology} \textbf{2018}, \emph{36}, 16
  3427.

\bibitem{dynes2016ultra}
J.~F. Dynes, W.~W. Tam, A.~Plews, B.~Fr{\"o}hlich, A.~W. Sharpe, M.~Lucamarini,
  Z.~Yuan, C.~Radig, A.~Straw, T.~Edwards, A.~J. Shields,
\newblock \emph{Scientific Reports} \textbf{2016}, \emph{6}.

\bibitem{Xu2015}
F.~Xu, M.~Curty, B.~Qi, L.~Qian, H.-K. Lo,
\newblock \emph{Nature Photonics} \textbf{2015}, \emph{9}, 12 772.

\bibitem{pirandolaPRL09}
S.~Pirandola, R.~Garc\'{\i}a-Patr\'on, S.~L. Braunstein, S.~Lloyd,
\newblock \emph{Phys. Rev. Lett.} \textbf{2009}, \emph{102} 050503.

\bibitem{tgw14}
M.~Takeoka, S.~Guha, M.~M. Wilde,
\newblock \emph{Nature Communications} \textbf{2014}, \emph{5}, 1 5235.

\bibitem{Pirandola2017}
S.~Pirandola, R.~Laurenza, C.~Ottaviani, L.~Banchi,
\newblock \emph{Nature Communications} \textbf{2017}, \emph{8}, 1 15043.

\bibitem{memory2011RMP}
N.~Sangouard, C.~Simon, H.~de~Riedmatten, N.~Gisin,
\newblock \emph{Rev. Mod. Phys.} \textbf{2011}, \emph{83} 33.

\bibitem{peev2009secoqc}
M.~Peev, C.~Pacher, R.~Alléaume, C.~Barreiro, J.~Bouda, W.~Boxleitner,
  T.~Debuisschert, E.~Diamanti, M.~Dianati, J.~F. Dynes, S.~Fasel, S.~Fossier,
  M.~Fürst, J.-D. Gautier, O.~Gay, N.~Gisin, P.~Grangier, A.~Happe, Y.~Hasani,
  M.~Hentschel, H.~Hübel, G.~Humer, T.~Länger, M.~Legré, R.~Lieger,
  J.~Lodewyck, T.~Lorünser, N.~Lütkenhaus, A.~Marhold, T.~Matyus,
  O.~Maurhart, L.~Monat, S.~Nauerth, J.-B. Page, A.~Poppe, E.~Querasser,
  G.~Ribordy, S.~Robyr, L.~Salvail, A.~W. Sharpe, A.~J. Shields, D.~Stucki,
  M.~Suda, C.~Tamas, T.~Themel, R.~T. Thew, Y.~Thoma, A.~Treiber, P.~Trinkler,
  R.~Tualle-Brouri, F.~Vannel, N.~Walenta, H.~Weier, H.~Weinfurter,
  I.~Wimberger, Z.~L. Yuan, H.~Zbinden, A.~Zeilinger,
\newblock \emph{New Journal of Physics} \textbf{2009}, \emph{11}, 7 075001.

\bibitem{sasaki2011field}
M.~Sasaki, M.~Fujiwara, H.~Ishizuka, W.~Klaus, K.~Wakui, M.~Takeoka, S.~Miki,
  T.~Yamashita, Z.~Wang, A.~Tanaka, K.~Yoshino, Y.~Nambu, S.~Takahashi,
  A.~Tajima, A.~Tomita, T.~Domeki, T.~Hasegawa, Y.~Sakai, H.~Kobayashi,
  T.~Asai, K.~Shimizu, T.~Tokura, T.~Tsurumaru, M.~Matsui, T.~Honjo, K.~Tamaki,
  H.~Takesue, Y.~Tokura, J.~F. Dynes, A.~R. Dixon, A.~W. Sharpe, Z.~L. Yuan,
  A.~J. Shields, S.~Uchikoga, M.~Legr\'{e}, S.~Robyr, P.~Trinkler, L.~Monat,
  J.-B. Page, G.~Ribordy, A.~Poppe, A.~Allacher, O.~Maurhart, T.~L\"{a}nger,
  M.~Peev, A.~Zeilinger,
\newblock \emph{Optics Express} \textbf{2011}, \emph{19}, 11 10387.

\bibitem{Zhang:18}
Q.~Zhang, F.~Xu, Y.-A. Chen, C.-Z. Peng, J.-W. Pan,
\newblock \emph{Opt. Express} \textbf{2018}, \emph{26}, 18 24260.

\bibitem{Qi07}
B.~Qi, C.-H.~F. Fung, H.-K. Lo, X.~Ma,
\newblock \emph{Quantum Info. Comput.} \textbf{2007}, \emph{7}, 1 73.

\bibitem{Lydersen2010}
L.~Lydersen, C.~Wiechers, C.~Wittmann, D.~Elser, J.~Skaar, V.~Makarov,
\newblock \emph{Nature Photonics} \textbf{2010}, \emph{4}, 10 686.

\bibitem{Xu_2010njp}
F.~Xu, B.~Qi, H.-K. Lo,
\newblock \emph{New Journal of Physics} \textbf{2010}, \emph{12}, 11 113026.

\bibitem{Lo2014}
H.-K. Lo, M.~Curty, K.~Tamaki,
\newblock \emph{Nature Photonics} \textbf{2014}, \emph{8} 595.

\bibitem{Lucamarini2018}
M.~Lucamarini, Z.~L. Yuan, J.~F. Dynes, A.~J. Shields,
\newblock \emph{Nature} \textbf{2018}, \emph{557}, 7705 400.

\bibitem{ma18}
X.~Ma, P.~Zeng, H.~Zhou,
\newblock \emph{Phys. Rev. X} \textbf{2018}, \emph{8} 031043.

\bibitem{tamaki2018information}
K.~Tamaki, H.-K. Lo, W.~Wang, M.~Lucamarini,
\newblock \emph{arXiv preprint arXiv:1805.05511} \textbf{2018}.

\bibitem{snsqkd}
X.-B. Wang, Z.-W. Yu, X.-L. Hu,
\newblock \emph{Phys. Rev. A} \textbf{2018}, \emph{98} 062323.

\bibitem{cui19}
C.~Cui, Z.-Q. Yin, R.~Wang, W.~Chen, S.~Wang, G.-C. Guo, Z.-F. Han,
\newblock \emph{Phys. Rev. Applied} \textbf{2019}, \emph{11} 034053.

\bibitem{Curty2019}
M.~Curty, K.~Azuma, H.-K. Lo,
\newblock \emph{npj Quantum Information} \textbf{2019}, \emph{5}, 1 64.

\bibitem{lin18}
J.~Lin, N.~L\"utkenhaus,
\newblock \emph{Phys. Rev. A} \textbf{2018}, \emph{98} 042332.

\bibitem{Yu2019}
Z.-W. Yu, X.-L. Hu, C.~Jiang, H.~Xu, X.-B. Wang,
\newblock \emph{Scientific Reports} \textbf{2019}, \emph{9}, 1 3080.

\bibitem{Zhang:19tfqkd}
C.-H. Zhang, C.-M. Zhang, Q.~Wang,
\newblock \emph{Opt. Lett.} \textbf{2019}, \emph{44}, 6 1468.

\bibitem{lu2019tfqkd}
F.-Y. Lu, Z.-Q. Yin, C.-H. Cui, G.-J. Fan-Yuan, R.~Wang, S.~Wang, W.~Chen,
  D.-Y. He, W.~Huang, B.-J. Xu, G.-C. Guo, Z.-F. Han,
\newblock \emph{Phys. Rev. A} \textbf{2019}, \emph{100} 022306.

\bibitem{Lu_2019}
F.-Y. Lu, Z.-Q. Yin, R.~Wang, G.-J. Fan-Yuan, S.~Wang, D.-Y. He, W.~Chen,
  W.~Huang, B.-J. Xu, G.-C. Guo, Z.-F. Han,
\newblock \emph{New Journal of Physics} \textbf{2019}, \emph{21}, 12 123030.

\bibitem{Yin2019}
H.-L. Yin, Z.-B. Chen,
\newblock \emph{Scientific Reports} \textbf{2019}, \emph{9}, 1 14918.

\bibitem{zhou2019PRA}
X.-Y. Zhou, C.-H. Zhang, C.-M. Zhang, Q.~Wang,
\newblock \emph{Phys. Rev. A} \textbf{2019}, \emph{99} 062316.

\bibitem{Grasselli2019}
F.~Grasselli, M.~Curty,
\newblock \emph{New Journal of Physics} \textbf{2019}, \emph{21}, 7 073001.

\bibitem{Maeda2019}
K.~Maeda, T.~Sasaki, M.~Koashi,
\newblock \emph{Nature Communications} \textbf{2019}, \emph{10}, 1 3140.

\bibitem{Yin2019SR}
H.-L. Yin, Y.~Fu,
\newblock \emph{Scientific Reports} \textbf{2019}, \emph{9}, 1 3045.

\bibitem{Yin2019a}
H.-L. Yin, Z.-B. Chen,
\newblock \emph{Scientific Reports} \textbf{2019}, \emph{9}, 1 17113.

\bibitem{wang2019twinfield}
R.~Wang, Z.-Q. Yin, F.-Y. Lu, S.~Wang, W.~Chen, W.~Huang, B.-J. Xu, G.-C. Guo,
  Z.-F. Han,
\newblock \emph{arXiv preprint arXiv:1904.07074} \textbf{2019}.

\bibitem{Grasselli2019asymm}
F.~Grasselli, {\'{A}}.~Navarrete, M.~Curty,
\newblock \emph{New Journal of Physics} \textbf{2019}, \emph{21}, 11 113032.

\bibitem{Wang_2020}
W.~Wang, H.-K. Lo,
\newblock \emph{New Journal of Physics} \textbf{2020}, \emph{22}, 1 013020.

\bibitem{Li2019}
W.~Li, L.~Wang, S.~Zhao,
\newblock \emph{Scientific Reports} \textbf{2019}, \emph{9}, 1 15466.

\bibitem{hu2019PRA}
X.-L. Hu, C.~Jiang, Z.-W. Yu, X.-B. Wang,
\newblock \emph{Phys. Rev. A} \textbf{2019}, \emph{100} 062337.

\bibitem{jiang2019zigzag}
C.~Jiang, X.-L. Hu, H.~Xu, Z.-W. Yu, X.-B. Wang,
\newblock \emph{arXiv preprint arXiv:1908.05670} \textbf{2019}.

\bibitem{lorenzo2019tight}
G.~C. Lorenzo, A.~Navarrete, K.~Azuma, G.~Kato, M.~Curty, M.~Razavi,
\newblock \emph{arXiv preprint arXiv:1910.11407} \textbf{2019}.

\bibitem{Minder2019}
M.~Minder, M.~Pittaluga, G.~L. Roberts, M.~Lucamarini, J.~F. Dynes, Z.~L. Yuan,
  A.~J. Shields,
\newblock \emph{Nature Photonics} \textbf{2019}, \emph{13}, 5 334.

\bibitem{wang19}
S.~Wang, D.-Y. He, Z.-Q. Yin, F.-Y. Lu, C.-H. Cui, W.~Chen, Z.~Zhou, G.-C. Guo,
  Z.-F. Han,
\newblock \emph{Phys. Rev. X} \textbf{2019}, \emph{9} 021046.

\bibitem{liu19}
Y.~Liu, Z.-W. Yu, W.~Zhang, J.-Y. Guan, J.-P. Chen, C.~Zhang, X.-L. Hu, H.~Li,
  C.~Jiang, J.~Lin, T.-Y. Chen, L.~You, Z.~Wang, X.-B. Wang, Q.~Zhang, J.-W.
  Pan,
\newblock \emph{Phys. Rev. Lett.} \textbf{2019}, \emph{123} 100505.

\bibitem{zhong19}
X.~Zhong, J.~Hu, M.~Curty, L.~Qian, H.-K. Lo,
\newblock \emph{Phys. Rev. Lett.} \textbf{2019}, \emph{123} 100506.

\bibitem{Fang2020}
X.-T. Fang, P.~Zeng, H.~Liu, M.~Zou, W.~Wu, Y.-L. Tang, Y.-J. Sheng, Y.~Xiang,
  W.~Zhang, H.~Li, Z.~Wang, L.~You, M.-J. Li, H.~Chen, Y.-A. Chen, Q.~Zhang,
  C.-Z. Peng, X.~Ma, T.-Y. Chen, J.-W. Pan,
\newblock \emph{Nature Photonics} \textbf{2020}.

\bibitem{chen509snstfqkd}
J.-P. Chen, C.~Zhang, Y.~Liu, C.~Jiang, W.~Zhang, X.-L. Hu, J.-Y. Guan, Z.-W.
  Yu, H.~Xu, J.~Lin, M.-J. Li, H.~Chen, H.~Li, L.~You, Z.~Wang, X.-B. Wang,
  Q.~Zhang, J.-W. Pan,
\newblock \emph{Phys. Rev. Lett.} \textbf{2020}, \emph{124} 070501.

\bibitem{zhong2020proof}
X.~Zhong, W.~Wang, L.~Qian, H.-K. Lo,
\newblock \emph{arXiv preprint arXiv:2001.10599} \textbf{2020}.

\bibitem{Pirandola2019}
S.~Pirandola,
\newblock \emph{Communications Physics} \textbf{2019}, \emph{2}, 1 51.

\bibitem{tamaki2012PRA}
K.~Tamaki, H.-K. Lo, C.-H.~F. Fung, B.~Qi,
\newblock \emph{Phys. Rev. A} \textbf{2012}, \emph{85} 042307.

\bibitem{hwang2003quantum}
W.-Y. Hwang,
\newblock \emph{Phys. Rev. Lett.} \textbf{2003}, \emph{91}, 5 057901.

\bibitem{wang2005beating}
X.-B. Wang,
\newblock \emph{Phys. Rev. Lett.} \textbf{2005}, \emph{94}, 23 230503.

\bibitem{lo2005decoy}
H.-K. Lo, X.~Ma, K.~Chen,
\newblock \emph{Phys. Rev. Lett.} \textbf{2005}, \emph{94}, 23 230504.

\bibitem{ma2012}
X.~Ma, M.~Razavi,
\newblock \emph{Phys. Rev. A} \textbf{2012}, \emph{86} 062319.

\bibitem{ferenczi2013security}
A.~Ferenczi,
\newblock \emph{Security proof methods for quantum key distribution protocols},
\newblock University of Waterloo, \textbf{2013}.

\bibitem{townsend93}
P.~D. {Townsend}, J.~G. {Rarity}, P.~R. {Tapster},
\newblock \emph{Electronics Letters} \textbf{1993}, \emph{29}, 7 634.

\bibitem{GLLP04}
D.~Gottesman, H.-K. Lo, N.~L\"{u}tkenhaus, J.~Preskill,
\newblock \emph{Quantum Info. Comput.} \textbf{2004}, \emph{4}, 5 325.

\bibitem{Lo2000science}
H.-K. Lo, H.~F. Chau,
\newblock \emph{Science} \textbf{1999}, \emph{283}, 5410 2050.

\bibitem{shor00}
P.~W. Shor, J.~Preskill,
\newblock \emph{Phys. Rev. Lett.} \textbf{2000}, \emph{85} 441.

\bibitem{DWbound}
I.~Devetak, A.~Winter,
\newblock \emph{Proceedings of the Royal Society A: Mathematical, Physical and
  Engineering Sciences} \textbf{2005}, \emph{461}, 2053 207.

\bibitem{losstolerant}
K.~Tamaki, M.~Curty, G.~Kato, H.-K. Lo, K.~Azuma,
\newblock \emph{Phys. Rev. A} \textbf{2014}, \emph{90} 052314.

\bibitem{Duan2001}
L.-M. Duan, M.~D. Lukin, J.~I. Cirac, P.~Zoller,
\newblock \emph{Nature} \textbf{2001}, \emph{414}, 6862 413.

\bibitem{biham1996}
E.~Biham, B.~Huttner, T.~Mor,
\newblock \emph{Phys. Rev. A} \textbf{1996}, \emph{54} 2651.

\bibitem{Inamori2002}
Inamori,
\newblock \emph{Algorithmica} \textbf{2002}, \emph{34}, 4 340.

\bibitem{aopp}
H.~Xu, Z.-W. Yu, C.~Jiang, X.-L. Hu, X.-B. Wang,
\newblock \emph{Phys. Rev. A} \textbf{2020}, \emph{101} 042330.

\bibitem{jiang2019}
C.~Jiang, Z.-W. Yu, X.-L. Hu, X.-B. Wang,
\newblock \emph{Phys. Rev. Applied} \textbf{2019}, \emph{12} 024061.

\bibitem{zeng2020}
P.~Zeng, W.~Wu, X.~Ma,
\newblock \emph{Phys. Rev. Applied} \textbf{2020}, \emph{13} 064013.

\bibitem{wangPRX19}
W.~Wang, F.~Xu, H.-K. Lo,
\newblock \emph{Phys. Rev. X} \textbf{2019}, \emph{9} 041012.

\bibitem{liu19mdi}
H.~Liu, W.~Wang, K.~Wei, X.-T. Fang, L.~Li, N.-L. Liu, H.~Liang, S.-J. Zhang,
  W.~Zhang, H.~Li, L.~You, Z.~Wang, H.-K. Lo, T.-Y. Chen, F.~Xu, J.-W. Pan,
\newblock \emph{Phys. Rev. Lett.} \textbf{2019}, \emph{122} 160501.

\bibitem{Ferrero:08}
V.~Ferrero, S.~Camatel,
\newblock \emph{Opt. Express} \textbf{2008}, \emph{16}, 2 818.

\bibitem{pound1946}
R.~V. Pound,
\newblock \emph{Review of Scientific Instruments} \textbf{1946}, \emph{17}, 11
  490.

\bibitem{Drever1983}
R.~W.~P. Drever, J.~L. Hall, F.~V. Kowalski, J.~Hough, G.~M. Ford, A.~J.
  Munley, H.~Ward,
\newblock \emph{Applied Physics B} \textbf{1983}, \emph{31}, 2 97.

\bibitem{liu2020injection}
Z.~{Liu}, R.~{Slavík},
\newblock \emph{Journal of Lightwave Technology} \textbf{2020}, \emph{38}, 1
  43.

\bibitem{Comandar2016}
L.~C. Comandar, M.~Lucamarini, B.~Fröhlich, J.~F. Dynes, A.~W. Sharpe,
  S.~W.-B. Tam, Z.~L. Yuan, R.~V. Penty, A.~J. Shields,
\newblock \emph{Nature Photonics} \textbf{2016}, \emph{10}, 5 312.

\bibitem{ma1995spie}
L.-S. Ma, P.~Jungner, J.~Ye, J.~L. Hall,
\newblock In Y.~Shevy, editor, \emph{Laser Frequency Stabilization and Noise
  Reduction}, volume 2378. International Society for Optics and Photonics,
  SPIE, \textbf{1995} 165 -- 175,
\newblock \urlprefix\url{https://doi.org/10.1117/12.208231}.

\bibitem{plugnplay}
A.~Muller, T.~Herzog, B.~Huttner, W.~Tittel, H.~Zbinden, N.~Gisin,
\newblock \emph{Applied Physics Letters} \textbf{1997}, \emph{70}, 7 793.

\bibitem{kerman2006}
A.~J. Kerman, E.~A. Dauler, W.~E. Keicher, J.~K.~W. Yang, K.~K. Berggren,
  G.~Gol’tsman, B.~Voronov,
\newblock \emph{Applied Physics Letters} \textbf{2006}, \emph{88}, 11 111116.

\bibitem{lv2018snspd}
C.~Lv, W.~Zhang, L.~You, P.~Hu, H.~Wang, H.~Li, C.~Zhang, J.~Huang, Y.~Wang,
  X.~Yang, Z.~Wang, X.~Xie,
\newblock \emph{AIP Advances} \textbf{2018}, \emph{8}, 10 105018.

\bibitem{annunziata2010}
A.~J. Annunziata, O.~Quaranta, D.~F. Santavicca, A.~Casaburi, L.~Frunzio,
  M.~Ejrnaes, M.~J. Rooks, R.~Cristiano, S.~Pagano, A.~Frydman, D.~E. Prober,
\newblock \emph{Journal of Applied Physics} \textbf{2010}, \emph{108}, 8
  084507.

\bibitem{Liu_2012}
D.-K. Liu, S.-J. Chen, L.-X. You, Y.-L. Wang, S.~Miki, Z.~Wang, X.-M. Xie,
  M.-H. Jiang,
\newblock \emph{Applied Physics Express} \textbf{2012}, \emph{5}, 12 125202.

\bibitem{agrawal2007nonlinear}
G.~P. Agrawal,
\newblock \emph{Nonlinear Fiber Optics},
\newblock Academic press, \textbf{2007}.

\bibitem{Fortier2019}
T.~Fortier, E.~Baumann,
\newblock \emph{Communications Physics} \textbf{2019}, \emph{2}, 1 153.

\bibitem{Mao2018QKD}
Y.~Mao, B.-X. Wang, C.~Zhao, G.~Wang, R.~Wang, H.~Wang, F.~Zhou, J.~Nie,
  Q.~Chen, Y.~Zhao, Q.~Zhang, J.~Zhang, T.-Y. Chen, J.-W. Pan,
\newblock \emph{Opt. Express} \textbf{2018}, \emph{26}, 5 6010.

\bibitem{Pereira2019}
M.~Pereira, M.~Curty, K.~Tamaki,
\newblock \emph{npj Quantum Information} \textbf{2019}, \emph{5}, 1 62.

\bibitem{tang2016measurement}
Y.-L. Tang, H.-L. Yin, Q.~Zhao, H.~Liu, X.-X. Sun, M.-Q. Huang, W.-J. Zhang,
  S.-J. Chen, L.~Zhang, L.-X. You, Z.~Wang, Y.~Liu, C.-Y. Lu, X.~Jiang, X.~Ma,
  Q.~Zhang, T.-Y. Chen, J.-W. Pan,
\newblock \emph{Physical Review X} \textbf{2016}, \emph{6}, 1 011024.

\bibitem{sibson2017chip}
P.~Sibson, C.~Erven, M.~Godfrey, S.~Miki, T.~Yamashita, M.~Fujiwara, M.~Sasaki,
  H.~Terai, M.~G. Tanner, C.~M. Natarajan, et~al.,
\newblock \emph{Nat. Commun.} \textbf{2017}, \emph{8} 13984.

\bibitem{wei2019high}
K.~Wei, W.~Li, H.~Tan, Y.~Li, H.~Min, W.-J. Zhang, H.~Li, L.~You, Z.~Wang,
  X.~Jiang, et~al.,
\newblock \emph{arXiv preprint arXiv:1911.00690} \textbf{2019}.

\bibitem{zhao2020}
S.~Zhao, P.~Zeng, W.-F. Cao, X.-Y. Xu, Y.-Z. Zhen, X.~Ma, L.~Li, N.-L. Liu,
  K.~Chen,
\newblock \emph{Phys. Rev. Applied} \textbf{2020}, \emph{14} 024010.

\end{thebibliography}

\end{document}